\documentclass[preprint,preprintnumbers,amsmath,amssymb,floatfix,nofootinbib]{revtex4}
\usepackage{epsf}
\usepackage{graphicx}
\usepackage{subfigure}
\usepackage{bbold}
\usepackage{hyperref}
\usepackage{slashed}
\usepackage{mathtools}

\newcommand{\bea}{\begin{eqnarray}}
\newcommand{\eea}{\end{eqnarray}}

\def\beq{\begin{equation}}
\def\eeq{\end{equation}}

\newcommand{\lsim}{\lesssim}

\newcommand{\gsim}{\gtrsim}

\begin{document}

\title{Top Partners and Higgs Boson Production}
\author{Chien-Yi~Chen}
%\email[]{dawson@bnl.gov}
\author{S.~Dawson}
\author{I.~M.~Lewis}

\affiliation{
Department of Physics, Brookhaven National Laboratory, Upton, New York, 11973
\vspace*{.5in}}

\date{\today}

\begin{abstract}
The Higgs boson is produced  at the LHC through gluon fusion at roughly the Standard Model rate.  New colored
fermions, which can contribute to $gg\rightarrow h$, must have vector-like interactions in
order not to be in conflict with the experimentally measured rate.  We examine the size of the corrections to single
and double Higgs production from heavy vector-like fermions in $SU(2)_L$
singlets and doublets and search for regions of parameter space where double Higgs production is enhanced
relative to the Standard Model prediction.  We compare production rates and distributions for double
Higgs production from gluon fusion using an exact calculation, the low energy theorem (LET), where
the top quark and the heavy vector-like fermions are taken to be infinitely massive, and an effective theory (EFT) where  top
mass effects are included exactly and the effects of the heavy fermions are included to ${\cal O}\biggl({1\over M^2_X}\biggr)$.  Unlike the LET, the EFT gives an extremely accurate description of the kinematic distributions for double Higgs production.  
\end{abstract}

\maketitle

\section{Introduction}
Having discovered a particle with the generic properties of the Standard Model Higgs boson, the next
important step is to determine what, if any, deviations from the standard picture are allowed by the data.  The
observed production and decay modes of the Higgs boson are within $\sim 20\%$ of the expectation
for a weakly coupled Higgs particle\cite{Aad:2012tfa,Chatrchyan:2012ufa}
 and so the possibilities for new physics in the Higgs sector are
 highly constrained\cite{ATLAS-CONF-2014-010}.  
A convenient framework 
to examine possible new high scale physics 
 is 
the language of effective field theories, where the theory is constructed to reduce
to the Standard Model at the electroweak scale, but new interactions are allowed at higher scales.  We study
an extension of the Standard Model where there are new massive quarks
which are allowed to interact with the Standard Model particles, and thus 
potentially
modify Higgs production
and decay rates.
Heavy fermions occur in many Beyond the Standard Model (BSM)  scenarios, in particular 
Little Higgs models \cite{Berger:2012ec,Csaki:2003si,Perelstein:2003wd,Chen:2003fm} and composite Higgs
models\cite{Contino:2006qr,Agashe:2005dk,Agashe:2004rs,Giudice:2007fh,Azatov:2011qy} 
in which the Higgs is 
strongly interacting at high scales. 
%insert
Direct searches for the heavy fermions have been extensively studied in the literature \cite{Han:2005ru,Han:2008gy,Chen:2012uw,Aad:2012bt,Aad:2011yn,
ATLAS-CONF-2013-018,CMS:2014rda,CMS-PAS-B2G-12-015,Chatrchyan:2012af,Chatrchyan:2012fp,Chatrchyan:2012vu,CMS:2012ab,Chatrchyan:2011ay,Beauceron:2014ila}.
  We consider models with both charge ${2\over 3}$ and $-{1\over 3}$ heavy quarks, but
where there are no additional Higgs bosons beyond the Standard Model $SU(2)_L$ doublet. 

New heavy colored fermions which couple to the Higgs boson
 cannot occur in chiral multiplets  since they would give large
contributions to the rate for Higgs production from gluon fusion\cite{Anastasiou:2011qw,Kribs:2007nz}.
 A single $SU(2)_L$ heavy quark doublet 
with corresponding right-handed heavy quark singlets would increase
the gluon fusion Higgs production
rate by a factor $\sim 9$, which is definitively excluded.   Vector-like quarks, on the other hand,
decouple at high energy and can be accommodated both by precision electroweak data, and by Higgs production
measurements.  Models with a single multiplet
of new vector-like 
fermions have been studied extensively in the context of single and double Higgs production from
gluon fusion\cite{Dawson:2012mk,Dawson:2012di,Aguilar-Saavedra:2013qpa,Gillioz:2012se,Berger:2012ec,Fajfer:2013wca,Ellis:2014dza,Grober:2010yv,Gillioz:2013pba}.  The rates for both single and double Higgs production in this class of models are close to those
of the Standard Model and 
the gluon fusion processes
are insensitive to the top partner masses and couplings.   This general feature is a result of the structure of the quark mass matrix and can be
proven using the Higgs low energy theorems (LETs)\cite{Low:2009di,Azatov:2011qy,Delaunay:2013iia}.

We study  more complicated models with several multiplets of 
vector-like quarks in both $SU(2)_L$ doublet and singlet representations\cite{Cynolter:2008ea}, which are
allowed to mix with the Standard Model quarks and with each other.     Higgs production from gluon fusion
can be  significantly altered from the Standard Model prediction
 when this mixing is allowed\cite{Gillioz:2012se,Delaunay:2013iia,Grober:2010yv}.  We explore the possibility of having the double
Higgs production rate be strongly enhanced or suppressed relative to the Standard Model, while
keeping the single Higgs rate close to that of the Standard Model. 
Models  with multiple representations of vector-like fermions have also been considered
in the context of flavor, where they have been used to generate a hierarchy of masses for the Standard
Model fermions\cite{Agashe:2009di,Buras:2011ph}.

Effective field theory (EFT) techniques can be used  to integrate out the effects of  heavy fermions.  Low energy physics is then described
by an effective Lagrangian,
\begin{equation}
L_{eff}=L_{SM}+\sum_i
{f_iO_i\over \Lambda^2}+ . . . \, ,
\label{leff}
\end{equation}
where $O_i$ are the dimension$-6$ operators corresponding to new physics at the scale $\Lambda$. 
These operators have been catalogued under various assumptions\cite{Buchmuller:1985jz,Contino:2013kra,Giudice:2007fh} and in this paper, we consider
only those operators affecting the gluon fusion production of Higgs bosons. 
We calculate the contributions to the $f_i$ obtained by integrating out heavy vector-like quarks 
in $SU(2)_L$ singlet and doublet representations, using
the equations of motion.
The new physics arising from the heavy vector-like  quarks   yields corrections to the 
Standard Model $SU(2)_L\times U(1)$ gauge couplings and  to the Yukawa couplings of the
light fermions. 

For arbitrary fermion mass matrices, we compute both single and double Higgs production from gluon
fusion.  As a by product of our calculation, 
we compare rates found by diagonalizing the mass matrices exactly, from the 
effective theory of Eq. \ref{leff} which contains terms of ${\cal O}\biggl({m_t^2\over \Lambda^2}
\biggr)$, and from the
low energy theorems, where $m_t\rightarrow \infty$ along with the new vector-like
quarks, in order to establish the numerical accuracy of
the various approximations.

Section \ref{modelsec} contains a brief description of the class of models studied here.  A description of single
and double Higgs production using the LET description and the EFT with
top and bottom quark mass effects included is given in Section \ref{leteftsec}.  Analytic results in an example with small
mixing between the Standard Model $3^{rd}$ generation quarks and the heavy quarks are given in Section
\ref{smallmat} in order to give an intuitive understanding of the new physics resulting from integrating out the
heavy vector-like fermions, while Section \ref{secresults} summarizes limits from precision electroweak measurements. 
Our major results are contained in Section \ref{higgsres}, where total rates and distributions
for double Higgs production are given in the full theory, the LET, and the EFT.  Finally, some conclusions are
in Section \ref{concsec}. 

\section{The Model}
\label{modelsec}
We consider models where in addition to the Standard Model field content, there
are two vector-like $SU(2)_L$ singlets, $U$ and $D$, and one vector-like $SU(2)_L$ doublet, $Q$, with hypercharges $Y=4/3$, $-2/3$, and $1/3$, respectively. 
We only allow mixing between the new fermions and the $3^{rd}$ 
generation
Standard Model quarks since the interactions of the two light generations
of quarks are highly constrained. 
 The Standard Model $3^{rd}$ generation fermions are
\begin{equation}
q_L=\left(\begin{array}{c}
t_L\\
b_L\end{array}\right)
, t_R\, ,b_R\, ,
\end{equation}
and the heavy vector-like fermions are,
\begin{equation}
Q=\left(\begin{array}{c}
T\\
B\end{array}\right) 
\,,  U \, , D\, ,
\end{equation}
where the left- and right- handed components have identical transformation 
properties under $SU(2)_L\times U(1)$, allowing for Dirac mass terms.
Finally, the Higgs doublet takes its usual form in unitary gauge after electroweak symmetry breaking,
\begin{equation}
H=\left( 
\begin{array}{c} 
0\\
{v+h\over\sqrt{2}}\end{array}
\right)\, ,
\end{equation}
where $v=246$~GeV is the Higgs vacuum expectation value, and $h$ is the Higgs boson.  
The Standard Model Lagrangian involving the third generation fermions and the Higgs boson is,
\begin{eqnarray}
L_{SM}&=& i {\overline q}_L \slashed D q_L+i {\overline t}_R
\slashed Dt_R+i {\overline b}_R \slashed Db_R
-\biggl(\lambda_t{\overline q}_L{\tilde H} t_R+\lambda_b{\overline q}_L H b_R
+h.c.\biggr)+|D_\mu H|^2-V(H) ,\nonumber\\
\label{lsm}
\end{eqnarray}
where ${\tilde H}\equiv i \sigma_2 H^*$, $V(H)$ is the Higgs potential, $D_\mu=(\partial _\mu-i{g\over 2}T\cdot W_\mu
-i{g^\prime\over 2} Y B_\mu-ig_s t\cdot G_\mu)$, $T^a=\sigma^a$ for $SU(2)_L$ doublets, $T^a=0$ for $SU(2)_L$ singlets, $\sigma^a$ are the Pauli matrices, for the quarks $t$ are the $SU(3)_C$ fundamental representation matrices, for the Higgs $t=0$, and $Q=(Y+T^3)/2$ is the electric charge operator. 
The classical equations of motion corresponding to Eq. \ref{lsm} are \cite{Buchmuller:1985jz},
\begin{eqnarray}
i\slashed Dq_L&=& \lambda_t {\tilde H} t_R+\lambda_b H b_R \nonumber \\
i\slashed Dt_R&=& \lambda_t {\tilde H}^\dagger q_L\nonumber \\
i\slashed D b_R&=& \lambda_b H^\dagger q_L\, . 
\label{eomsm}
\end{eqnarray}

The most general Lagrangian coupling the third generation quarks and the new fermions is
$L_{NP}$,
\begin{eqnarray}
L_{NP}&\equiv & L^\prime_M+L^\prime_{KE}+L^\prime_{Y}\nonumber \\
L^\prime_M&=&-
M {\overline Q} Q-M_U {\overline U}U-M_D{\overline D}D
\nonumber \\ 
L^\prime_{KE}
&=& {\overline Q}(i\slashed D)Q+{\overline U}(i\slashed D)U+{\overline D}(i\slashed D)D
\nonumber \\ 
L^\prime_{Y}&=&
-\biggl\{
  \lambda_1 {\overline Q}_L{\tilde H} U_R
+\lambda_2 {\overline Q}_L H D_R
+\lambda_3 {\overline Q}_R{\tilde H}U_L
+M_4 {\overline q}_L Q_R
+M_5 {\overline U}_L t_R
+M_6 {\overline D}_L b_R
\nonumber \\ &&
+\lambda_7 {\overline q}_L {\tilde H} U_R+\lambda_8 {\overline q}_L H D_R
+\lambda_9 {\overline Q}_L{\tilde H}{ t_R}
+\lambda_{10} {\overline Q}_L H{ b_R}
+\lambda_{11} {\overline Q}_R H{ D_L}
+ h.c. \biggr\}\, .
\label{modeldef}
\end{eqnarray}

The much studied cases where the Standard Model top quark mixes with only a singlet or doublet 
vector-like fermion\cite{Dawson:2012di,Aguilar-Saavedra:2013qpa,delAguila:2000aa,delAguila:2000rc,Lavoura:1992np,Dawson:2012mk,Fajfer:2013wca,Lavoura:1992qd,AguilarSaavedra:2002kr} can be obtained from this study, as can the composite model case where the Standard Model
quarks do not couple to the Higgs doublet ($\lambda_t=\lambda_b=\lambda_7=\lambda_9=\lambda_{10}=0$).   We will consider various mass hierarchies
in the following sections.

The  mass  and Yukawa interactions can be written as
\begin{eqnarray}
- L_{Y^\prime}&=& 
{\overline \chi}^t_L  M^{(t)}(h) \chi_R^t+
{\overline \chi}^b_L M^{(b)}(h)\chi_R^b
+h.c.
\, ,
\end{eqnarray}
where $\chi^t_{L,R}\equiv(t,T,U)_{L,R}$, 
$\chi^b_{L,R}\equiv(b,B,D)_{L,R}$, and the Higgs-dependent fermion mass matrices are 

\begin{equation}
M^{(t)}(h)=\left(\begin{array}{c c c}
\lambda_t({h+v\over\sqrt{2}})&M_4&\lambda_7({h+v\over\sqrt{2}})\\
\lambda_9({h+v\over\sqrt{2}}) & M& \lambda_1({h+v\over\sqrt{2}})\\
M_5 & \lambda_3({h+v\over\sqrt{2}})& M_U
\end{array} 
\right)\, ,
\qquad 
M^{(b)}(h)=\left(\begin{array}{c c c}
\lambda_b({h+v\over\sqrt{2}})&M_4&\lambda_8({h+v\over\sqrt{2}})\\
\lambda_{10}({h+v\over\sqrt{2}}) & M& \lambda_2({h+v\over\sqrt{2}})\\
M_6 & \lambda_{11}({h+v\over\sqrt{2}})& M_D
\end{array} 
\right)\, ,
\label{massdef}
\end{equation}
where typically $\lambda_i\sim {\cal {O}}(1)$. 
The mass eigenstate fields, $\psi^t\equiv(T_1,T_2,T_3)$ and ${\psi^b\equiv(B_1,B_2,B_3)}$, are   found by means of bi-unitary
transformations,
\begin{eqnarray}
-L_{Y^\prime}&=& 
{\overline \chi}^t_L(V_L^{t\dagger} V_L^t) M^{(t)}(h)(V^{t\dagger}_R V^t_R) \chi_R^t+
{\overline \chi}^b_L (V_L^{b\dagger}V_L^b) M^{(b)}(h)(V_R^{b\dagger } V_R^b)\chi_R^b
+h.c.
\nonumber \\
&=& 
{\overline \psi}^t_L M_{diag}^t\psi^t_R
+{\overline \psi}^b_L M_{diag}^b\psi^b_R+{\overline \psi}^t_L {\cal Y}^t \psi^t_R h+{\overline \psi}^b_L {\cal Y}^b \psi^b_R h+h.c.
\, ,
\label{diags}
\end{eqnarray}
and $(T_1,B_1)$  are the Standard Model $3^{rd}$ generation quarks. 
The diagonal mass matrices can be written, 
\begin{eqnarray}
M_{diag}^t&=&V_L^tM^{(t)}(0) V_R^{t\dagger}\nonumber \\
(M_{diag}^{t})^2&=& 
V_L^t 
M^{(t)}(0) M^{(t)}(0)^{\dagger} V_L^{t\dagger}\nonumber \\
&=&
V_R^t 
M^{(t)}(0)^{\dagger}
M^{(t)}(0) V_R^{t \dagger}
\, ,
\label{yukints}
\end{eqnarray}
where we have set $h=0$, 
the Yukawa matrix is
\begin{eqnarray}
{\cal Y}^t\,h&=&V_L^t\left(M^{(t)}(h)-M^{(t)}(0)\right) V_R^{t\dagger}
\end{eqnarray}
and similarly in the $b$ sector. 

The couplings to the $W$ contain both left- and right- handed contributions,
\begin{eqnarray}
L_W&=& {g\over \sqrt{2}}\biggl(
{\overline \chi}_R^{t,2}
\gamma_\mu\chi_R^{b,2}+\sum_{i=1,2}
{\overline \chi}_L^{t,i} 
\gamma_\mu\chi_L^{b,i}\biggr)W^{+\mu}+h.c. \nonumber \\
&=& 
{g\over \sqrt{2}}\sum_{j,k=1,2,3}\biggl( {\overline \psi}_L^{t,j} (U_L)_{jk} \gamma_\mu\psi_L^{b,k}
+{\overline \psi}_R^{t,j}(U_R)_{jk}\gamma_\mu \psi_R^{b,k}\biggr)W^{+\mu} +h.c. \, ,
\label{wcoups}
\end{eqnarray}
where 
\begin{eqnarray}
(U_L)_{jk}&=& \sum_{i=1,2} (V^t_{L})_{ji} (V^{b\dagger}_L)_{ik}\nonumber \\
(U_R)_{jk}&=&  (V^t_{R})_{j2} (V^{b\dagger}_R)_{2k}
\label{wmatdef}
\end{eqnarray}

Finally, the couplings to the $Z$ are,
\begin{eqnarray}
L_Z&=&{g\over 2 c_W}
\sum_{j,k=1,2,3}
\biggl\{
{\overline \psi}^t_{L,j} (X_L^t)_{jk}\gamma_\mu \psi^t_{L,k}+
{\overline \psi}^t_{R,j} (X_R^t)_{jk}\gamma_\mu \psi^t_{R,k}
\nonumber \\
&&
-{\overline \psi}^b_{L,j} (X_L^b)_{jk}\gamma_\mu \psi^b_{L,k}
-{\overline \psi}^b_{R,j} (X_R^b)_{jk}\gamma_\mu \psi^b_{R,k}
\biggr\} Z^\mu
\nonumber \\
&&  - {g\over 2 c_W}(2s_W^2)J^\mu_{EM} Z_\mu
\, ,
\label{zcoup}
\end{eqnarray}
where $s_W=\sin\theta_W$, $c_W=\cos\theta_W$, $\theta_W$ is the weak mixing angle,
\begin{eqnarray}
(X_L^t)_{jk}&=& \sum_{i=1,2} (V^t_{L})_{ji} (V^{t\dagger}_L)_{ik}\nonumber \\
(X_R^t)_{jk}&=&  (V^t_{R})_{j2} (V^{t\dagger}_R)_{2k}
\nonumber\\ 
(X_L^b)_{jk}&=& \sum_{i=1,2} (V^b_{L})_{ji} (V^{b\dagger}_L)_{ik}\nonumber \\
(X_R^b)_{jk}&=&  (V^b_{R})_{j2} (V^{b\dagger}_R)_{2k}
\label{zcoupdef}
\end{eqnarray}
and $J_{EM}^{\mu}$ is the usual electromagnetic current,
\begin{eqnarray}
J^\mu_{EM}&=& Q_t\biggl[\overline{ \psi}^t_L\gamma^\mu \psi_L^t
+\overline{ \psi}^t_R\gamma^\mu \psi_R^t\biggr]
 +Q_b\biggl[\overline{ \psi}^b_L\gamma^\mu \psi_L^b
+\overline{ \psi}^b_R\gamma^\mu \psi_R^b\biggr]\, .
\end{eqnarray}
The $Z$ couplings contain flavor non-diagonal contributions due to the off diagonal terms
in $X_{L,R}^{t,b}$.  It is straightforward to apply the results of Eqs. \ref{wcoups} and \ref{zcoup} 
to find the gauge boson couplings in a specific model. 

\section{ Effective Theory Results}
\label{leteftsec}
In this section, we consider single and double Higgs production from gluon fusion in the general model
described in the previous section.   We begin with the results using the LET,
in which the top quark and all top partners are taken infinitely massive.  We next include
the top quark and bottom quark masses exactly and compute to ${\cal O}({1\over M_X^2})$, where $M_X$ is a
generic heavy vector fermion mass.  These results (EFT)  are then matched to an effective Lagrangian
to determine the coefficients of the dimension-6  operators.  We are interested
in comparing the numerical accuracy of the two approximations with the exact calculations for the gluon fusion rates.  
\subsection{Effective Theory From Low Energy Theorems}
The low energy theorems can be used to integrate out the effect of the charge ${2\over 3}$
 massive particles, 
including the top quark. 
In the limit in which fermion masses ($M_{T_1}, M_{T_2}, M_{T_3}$) are much heavier than the Higgs mass, the 
$hgg$ coupling can be found from the low energy effective interaction of a colored
Dirac  fermion with the gluon field strength\cite{Kniehl:1995tn},
\begin{equation}
L^{(t)}_{hgg}={\alpha_s\over 24 \pi}h
\biggl({\partial \over \partial
 h}\ln \biggl[\det
(M^{(t)}(h)^\dagger M^{(t)}(h))\biggr]\biggl)_{h=0}G^{A,\mu\nu}G^A_{\mu\nu}\, ,
\end{equation}
where $G^A_{\mu\nu}$ is the gluon field-strength tensor.  
With no approximation on the relative size of the 
parameters in $M^{(t)}$, the LET gives for the contributions
from the top sector alone,
\begin{equation}
L_{hgg}^{(t)}={\alpha_s\over 12 \pi}{h\over v}\biggl[
1+2\lambda_3 v^2 \biggl({\lambda_1\lambda_t-\lambda_7\lambda_9\over X}\biggr)
\biggr]G^{A,\mu\nu}G^A_{\mu\nu}
\, ,
\label{letggh}
\end{equation}
where,
\begin{equation}
X\equiv -\frac{v}{2\sqrt{2}}\det M^{(t)}(0)=v^2\lambda_3 (\lambda_1\lambda_t-\lambda_7\lambda_9) 
+2\biggl[
-\lambda_1M_4 M_5+M M_5\lambda_7 +M_UM_4\lambda_9-\lambda_t M M_U
\biggr] \, .
\end{equation}
Having non-zero $\lambda_3$, the coupling between the doublet and singlet vector-like quarks and Higgs boson, is critical for achieving a result which 
is different from the LET for the Standard Model:
\begin{equation}
L^{SM}_{h^ngg}=\frac{\alpha_s}{12\pi}\left[\frac{h}{v}-\frac{h^2}{2v^2}+\ldots\right]G^{A,\mu\nu}G^A_{\mu\nu}.
\label{comp}
\end{equation}
   This can be understood by noting that when the mass matrix factorizes,
\begin{eqnarray}
\det \left(M^{(t)}(h)\right)=F\biggl({h\over v}\biggr) G(\lambda_i,M_X,m_t)\, ,
\end{eqnarray}
the LET has no dependence on the 
heavy mass scales and Yukawa couplings
as in Eq. \ref{comp}~\cite{Low:2009di,Azatov:2011qy,Delaunay:2013iia}. In the limit $\lambda_3\rightarrow0$, we have:
\begin{equation}
\det M^{(t)}(h)\bigg|_{\lambda_3=0}=-\frac{h+v}{2\sqrt{2}} X\bigg|_{\lambda_3=0}.
\end{equation}
and the LET reduces to the Standard Model result.

In the limit $M, M_U \gg  M_5,M_4,v$ and all the Yukawa couplings $\lambda_i$ are ${\cal O}(1)$,
\begin{equation}
L_{hgg}^{(t)}\rightarrow {\alpha_s\over 12 \pi}{h\over v}\biggl[
1-\lambda_3 v^2 \biggl({\lambda_1\lambda_t-\lambda_7\lambda_9\over M M_U \lambda_t}\biggr)
\biggr]G^{\mu\nu, A}G^A_{\mu\nu}\, .
\label{EFTfull}
\end{equation}
If, motivated by composite models\cite{Azatov:2011qy}, 
we assume that there are no couplings of the Standard Model quarks
to the Higgs, then $\lambda_t=\lambda_b
=\lambda_7=\lambda_9=\lambda_{10}=0$, and  with no assumption about the
relative sizes of the remaining terms,
\begin{equation}
L_{hgg}^{(t)}\rightarrow {\alpha_s\over 12 \pi}{h\over v}
G^{A,\mu\nu}G^A_{\mu\nu}\, ,
%\label{comp}
\end{equation}
and the Standard Model result is recovered.   Similarly to above, in this limit the determinant of the mass matrix factorizes.

Double Higgs production can also be found 
using the LET~\cite{Kniehl:1995tn,Dawson:2012mk,Gillioz:2012se},  
 \begin{eqnarray}
L_{hhgg}^{(t)}&=&{\alpha_s\over 48 \pi}h^2
\biggl({\partial^2\over \partial h^2}\ln \biggl[\det M^{(t)}(h)^\dagger M^{(t)}(h) \biggr]\biggr)_{h=0}
G^{A,\mu\nu}G^A_{\mu\nu}\, ,
\end{eqnarray}
and we obtain,
\begin{eqnarray}
L_{hhgg}^{(t)}&=&-{\alpha_s\over 24\pi}{h^2\over v^2}
\biggl\{
1-2\lambda_3 v^2(\lambda_1\lambda_t-\lambda_7\lambda_9)
\biggl[
{1\over X}-{2\lambda_3 v^2(\lambda_1\lambda_t-\lambda_7\lambda_9)
\over X^2}\biggr] \biggr\}G^{A,\mu\nu}G^A_{\mu\nu}\, .
\label{letgghh}
\end{eqnarray}
In the limit $M,M_U \gg  M_4,M_5,v$,
\begin{eqnarray}
L_{hhgg}^{(t)}&=&-{\alpha_s\over 24\pi}{h^2\over v^2}
\biggl\{
1
+\lambda_3 v^2 \biggl({\lambda_1\lambda_t-\lambda_7\lambda_9\over M M_U \lambda_t}\biggr)
 \biggr\}G^{A,\mu\nu}G^A_{\mu\nu}\, .
\label{ggHHFull}
\end{eqnarray}

Since the $b$ quark is not a heavy fermion,  the effective $ggh$ Lagrangian in the charge $-{1\over 3}$ sector
requires more care and the LET cannot be naively applied.  In the next section, we formally integrate out the heavy $T_2,T_3$ and $B_2,B_3$ fields,
while retaining all mass dependence from the light Standard Model-like quarks, $T_1,B_1$.

\subsection{Effective Theory with Top and Bottom Quark Masses}

The effects of finite top and bottom quark masses can be included by using the classical
equations of motion to 
integrate out the heavy fields $T_2,T_3,B_2$ and $B_3$\cite{delAguila:2000aa,delAguila:2000rc,Buras:2011ph,Kilian:2003xt,Buchmuller:1985jz}.
We assume that $M, M_U$ and $M_D$ are
of similar magnitude and are much larger than $v$, that the Yukawa couplings, $\lambda_i$, are
of ${\cal O}(1)$,
and expand to ${\cal O}(1/M^2_X)$, 
%***** CHECK WHAT CHANGES WHEN $M_5$ is BIG *****
\begin{eqnarray}
U_L&=&\biggl(-{\lambda_7\over M_U}+{\lambda_1M_4\over M M_U}
-{\lambda_t M_5\over M_U^2}\biggr) ({\tilde H}^\dagger q_L)
+{\cal O}\biggl({1\over M^3_X}\biggr)
\nonumber \\
U_R&=& \biggr[-{M_5\over M_U}+
\biggl({\lambda_3\lambda_9\over M M_U}-
{\lambda_t \lambda_7\over M_U^2}\biggr)(H^\dagger H)\biggr] t_R
-{\lambda_7\over M_U^2} i(D_\mu {\tilde H})^\dagger\gamma^\mu q_L
+{\cal O}\biggl({1\over M^3_X}\biggr)
\nonumber \\
D_L&=&\biggl(-{\lambda_8\over M_D}+{\lambda_2 M_4\over M M_D}
-{\lambda_b M_6\over M_D^2}
\biggr) ( H^\dagger q_L)
+{\cal O}\biggl({1\over M^3_X}\biggr)
\nonumber \\
D_R&=& \biggr[-{M_6\over M_D}+
\biggl(
{\lambda_{10}\lambda_{11}\over M M_D}
-{\lambda_b\lambda_8\over M_D^2}\biggr)
(H^\dagger H)\biggr] b_R-{\lambda_8\over M_D^2} i(D_\mu H)^\dagger \gamma^\mu q_L
+{\cal O}\biggl({1\over M^3_X}\biggr)\nonumber \\
Q_L&=& \biggl[
-{M_4\over M}+
\biggl(
{\lambda_8\lambda_{11}\over M M_D}
-{\lambda_b\lambda_{10}\over M^2}
\biggr)(H H^\dagger)
+
\biggl(
{\lambda_3\lambda_7\over M M_U}
-{\lambda_t\lambda_9\over M^2}
\biggr)
{\tilde H} {\tilde H}^\dagger  
\biggr] q_L\nonumber \\
&&-{\lambda_9\over M^2}
(i\slashed D{\tilde H})t_R-{\lambda_{10}\over M^2}(i\slashed DH)b_R
+{\cal O}\biggl({1\over M^3_X}\biggr)\nonumber \\
Q_R&=& \biggl(-{\lambda_9\over M}+{\lambda_1 M_{5}\over M M_U}
-{\lambda_tM_4\over M^2}
\biggr){\tilde H}t_R
\nonumber \\ &&
+\biggl(-{\lambda_{10}\over M}+
{\lambda_2M_6\over M M_D}
-{\lambda_b M_4\over M^2}
\biggr)
Hb_R
+{\cal O}\biggl({1\over M^3_X}\biggr)
\, .
\nonumber \\
\label{heavyferms}
\end{eqnarray}
Substituting Eq. \ref{heavyferms} into $L^\prime_{M}+L^\prime_Y$,
\begin{eqnarray}
L^\prime_{M}+L^\prime_Y&\equiv&
L_{eff}^{(a)}\nonumber \\
&\rightarrow &
-\biggl\{
\biggl(-
{M_5\lambda_7\over M_U}
-{M_4\lambda_9\over M}
+{\lambda_1M_4 M_5\over MM_U}+{\lambda_3\lambda_7\lambda_9\over M M_U}(H^\dagger H)
\biggr){\overline q}_L {\tilde H}t_R
\nonumber \\ 
&&+\biggl(-{M_6\lambda_8\over M_D}-{M_4\lambda_{10}\over M}
+{\lambda_2 M_4M_6\over M M_D}
+{\lambda_8\lambda_{10}\lambda_{11}\over M M_D}(H^\dagger H)
\biggr)
{\overline q}_L Hb_R
\nonumber \\ &&
+h.c.\biggr\}+
\delta L_{eff}^h+
{\cal O}\biggl({1\over M^3_X}\biggr)\, ,
\label{yukeff}
\end{eqnarray}
where ${\delta L}_{eff}^h$ collects the contributions from the terms in Eq. \ref{heavyferms} containing
derivatives of the Higgs field\cite{delAguila:2000rc},
\begin{eqnarray}
\delta L_{eff}^h&=& \biggl\{
{1\over 4} \biggl({\lambda_7^2\over M_U^2}-{\lambda_8^2\over M_D^2}\biggr)
(H^\dagger iD_\mu H)
({\overline q}_L\gamma^\mu q_L)
\nonumber \\
&&
-{1\over 4} \biggl({\lambda_7^2\over M_U^2}+{\lambda_8^2\over M_D^2}\biggr)
(H^\dagger  \sigma^a iD_\mu H)
({\overline q}_L\gamma^\mu \sigma^a q_L)\nonumber \\
&&-{\lambda_9^2\over 2 M^2} \biggl[
(H^\dagger iD_\mu H)({\overline t}_R\gamma^\mu t_R)\biggr]
+{\lambda_{10}^2\over 2 M^2}\biggl[(H^\dagger iD_\mu H)
({\overline b}_R\gamma^\mu b_R)\biggr]
\nonumber \\ &&
+{\lambda_9\lambda_{10}\over M^2} \biggl[(
{\tilde H}^\dagger iD_\mu H) ( {\overline t}_R\gamma^\mu b_R)\biggr]\biggr\} +h.c.
+{\cal O}\biggl({1\over M^3_X}\biggr)\, .
\label{higgslf}
\end{eqnarray}
Eq. \ref{higgslf} corresponds to $\Delta L_{F_1}$ of Refs. \cite{Contino:2013kra,Alloul:2013naa}.

Similarly, substituting Eq. \ref{heavyferms} into the kinetic energy terms of $L^\prime_{KE}$,
\begin{eqnarray}
L^\prime_{KE}
&\equiv &
L_{eff}^{(b)}
\nonumber \\ 
&\rightarrow & {1\over 2}\biggl\{
 \lambda_t {\overline q}_L {\tilde H} t_R \biggl(
{M_4^2\over M^2}+{M_5^2\over M_U^2}
+(H^\dagger H)\biggl[ {\lambda_7^2\over M_U^2}
+{\lambda_9^2\over M^2}
\biggr]\biggr)
\nonumber \\
&& +\lambda_b {\overline q}_L { H} b_R \biggl(
{M_4^2\over M^2}+{M_6^2\over M_D^2}
+(H^\dagger H)\biggl[ {\lambda_8^2\over M_D^2}
+{\lambda_{10}^2\over M^2}
\biggr]\biggr)\biggr\} 
\nonumber \\ &&
+h.c. + {\cal{O}}\biggl({1\over M^3_X}\biggr)\, .
\end{eqnarray} 

\begin{figure}[tb]
\begin{center}
\subfigure[]{\includegraphics[width=0.8\textwidth,clip]{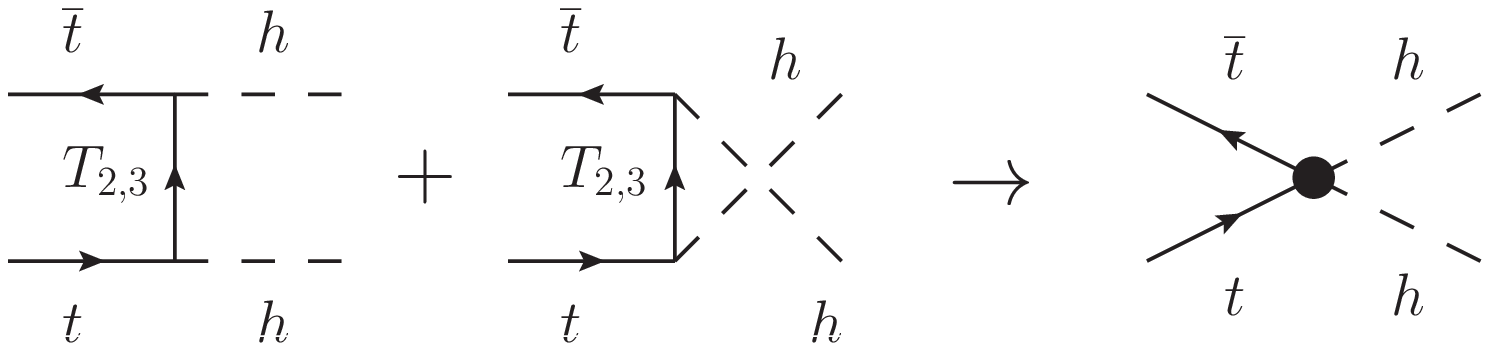}}
\subfigure[]{\includegraphics[width=0.9\textwidth,clip]{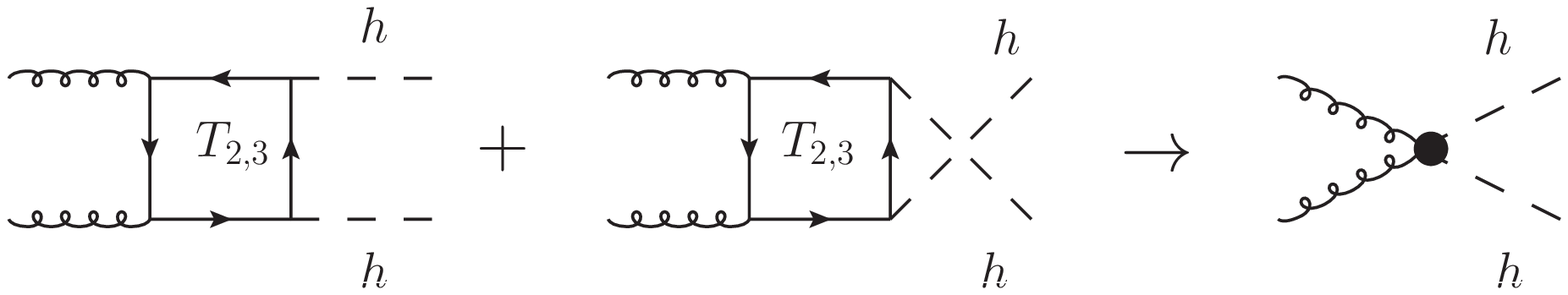}}
\subfigure[]{\includegraphics[width=0.6\textwidth,clip]{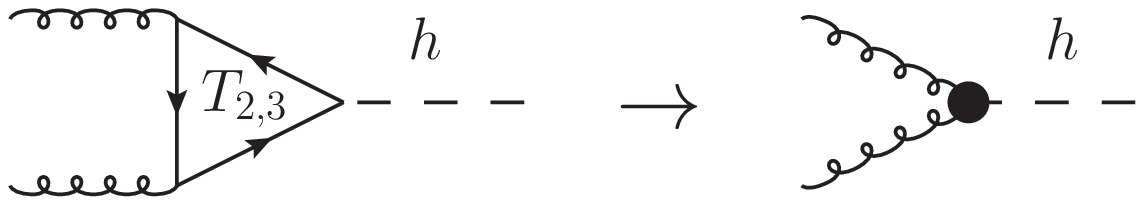}}
\end{center}
\caption{Representative diagrams corresponding to integrating out heavy fields and generating the (a) $\bar{t}thh$, (b) $G^{A,\mu\nu}G^{A}_{\mu\nu}h^2$, and (c) $G^{A,\mu\nu}G^{A}_{\mu\nu}h$ operators in Eq.~\ref{effective}.}
\label{fig:EFT}
\end{figure}
The effective  low energy Lagrangian after electroweak symmetry breaking contains only Standard
Model fields, but non-Standard Model coefficients and operators have been generated
by integrating out the heavy fields. The procedure of integrating out by the equations of motion occurs at tree level.  However, at loop level, integrating out heavy colored particles will generate operators of the form $G^{A,\mu\nu}G^{A}_{\mu\nu}h^2$ and  $G^{A,\mu\nu}G^{A}_{\mu\nu}h$, which need to be included in the effective Lagrangian:
\begin{eqnarray}
L_{eff}&=&L_{SM}+L_{eff}^{(a)}+L_{eff}^{(b)}+\delta L_{eff}^h+{c_g\alpha_s\over 12 \pi v}G^{A,\mu\nu}G_A^{\mu\nu}h
-{c_{gg}\alpha_s\over 24 \pi v^2}G^{A,\mu\nu}G_A^{\mu\nu}h^2\nonumber \\
 &=& i {\overline q}_L \slashed D q_L+i {\overline t}_R
 \slashed Dt_R+i {\overline b}_R \slashed Db_R+|D_\mu H|^2-V(H)
-m_t {\overline t}t -Y_t{\overline t} t h
+c_{2h}^{(t)}{\overline t}t h^2\nonumber \\ &&
-m_b{\overline b} b 
-Y_b{\overline b} b h
+c_{2h}^{(b)}{\overline b} b h^2+{c_g\alpha_s\over 12 \pi v}G^{A,\mu\nu}G_A^{\mu\nu}h
-{c_{gg}\alpha_s\over 24 \pi v^2}G^{A,\mu\nu}G_A^{\mu\nu}h^2\nonumber \\
&&+{g\over \sqrt{2}}\biggl\{\biggl[
\delta g_L\,{\overline t}_L\gamma^\mu b_L  
+\delta g_R\,{\overline t}_R\gamma^\mu b_R \biggr]W^+_\mu+h.c.\biggr\}
\nonumber \\
&&+{g\over c_W}\biggl\{ 
{\overline t}_L \gamma_\mu t_L\,
\delta Z_L^t
+{\overline t}_R\gamma_\mu t_R\,
\delta Z_R^t
%\nonumber \\
+{\overline b}_L\gamma_\mu b_L\,
\delta Z_L^b
+{\overline b}_R\gamma_\mu b_R\,
\delta Z_R^b
\biggr\} Z^\mu
\nonumber \\
&&+\delta {L^{h'}_{eff}}+{\cal O}\biggl({1\over M^3_X}\biggr) \, .
\label{effective}
\end{eqnarray}
The non-Standard Model like gauge boson coupling in lines 4 and 5 in the above equation originate from $\delta L_{eff}^h$, and $\delta L_{eff}^{h'}$ is defined to be $\delta L_{eff}^h$ with these terms removed.  
In Fig.~\ref{fig:EFT} we show representative diagrams illustrating the generation of the (a)~$\bar{t}th^2$, (b)~$G^{A,\mu\nu}G^{A}_{\mu\nu}h^2$, and (c)~$G^{A,\mu\nu}G^{A}_{\mu\nu}h$ effective operators.
To ${\cal O}\biggl(
{1\over M^2_X}\biggr)$,  the Yukawa couplings are shifted from their Standard Model values,
\begin{eqnarray}
\sqrt{2} Y_t&=&\lambda_t\biggl\{
1-\biggl[
{M_4^2\over 2 M^2}+
{M_5^2\over 2 M_U^2}
+{3v^2\over 4}
\biggl(
{\lambda_7^2\over M_U^2}
+{\lambda_9^2\over M^2}\biggr)
\biggr]
\biggr\}
\nonumber \\ 
&& -
{M_5\lambda_7\over M_U}
-{M_4\lambda_9\over M}
+{\lambda_1M_4 M_5\over MM_U}
+{3v^2\over 2 M M_U}\lambda_3\lambda_7\lambda_9
\nonumber \\
&=& \sqrt{2}{m_t\over v} +{v^2\over M M_U}\lambda_3\lambda_7\lambda_9
-\lambda_t {v^2\over 2} \biggl(
{\lambda_7^2\over M_U^2}+{\lambda_9^2\over M^2}\biggr)
\nonumber \\
{Y_tv\over m_t}&\equiv 1+\delta_t
\nonumber \\
\sqrt{2} Y_b&=&\lambda_b\biggl\{
1-\biggl[
{M_4^2\over 2 M^2}+
{M_6^2\over 2 M_D^2}
+{3v^2\over 4}
\biggl(
{\lambda_8^2\over M_D^2}
+{\lambda_{10}^2\over M^2}\biggr)
\biggr]
\biggr\}
\nonumber \\ 
&& -
{M_6\lambda_8\over M_D}
-{M_4\lambda_{10}\over M}
+{\lambda_2 M_4 M_6\over MM_D}
+{3v^2\over 2 M M_D}\lambda_8\lambda_{10}\lambda_{11}
\nonumber \\
&=& \sqrt{2}{m_b\over v} +{v^2\over M M_D}\lambda_8\lambda_{10}\lambda_{11}
-\lambda_b {v^2\over 2} \biggl(
{\lambda_8^2\over M_D^2}+{\lambda_{10}^2\over M^2}\biggr)
\nonumber \\
{Y_bv\over m_b}&\equiv 1+\delta_b\, .
\end{eqnarray}
We see that $Y_t$ and $Y_b$  are no longer proportional to $m_t=M_{T_1}$ and $m_b=M_{B_1}$.
Non-Standard Model couplings of the fermions to Higgs pairs are also generated, as 
are Higgs-gluon effective couplings,
\begin{eqnarray}
\nonumber \\
{c_{2h}^{(t)}\over v} &=& {3\over 2\sqrt{2}}
\biggl\{-{\lambda_3 \lambda_7\lambda_9\over M M_U}
+{1\over 2}
\lambda_t \biggl(
{\lambda_7^2\over M_U^2}+
{\lambda_9^2\over M^2}\biggr)\biggr\}
\nonumber \\
&=&-{3\over 2 v^2}\biggl(Y_t-{m_t\over v}\biggr)\nonumber \\
&=& -{3\over 2}{m_t\delta_t\over v^3}\nonumber \\ 
{c_{2h}^{(b)}\over v}&=& {3\over 2\sqrt{2}}
\biggl\{-{\lambda_8 \lambda_{10}\lambda_{11}\over M M_D}
+{1\over 2}
\lambda_b \biggl(
{\lambda_8^2\over M_D^2}+
{\lambda_{10}^2\over M^2}\biggr)\biggr\}
\nonumber \\
&=& -{3\over 2} {m_b \delta_b\over v^3} 
\nonumber \\
c_g&=& v^2\biggl[-{\lambda_1\lambda_3\over M M_U}
-{\lambda_2\lambda_{11}\over M M_D} 
+{1\over 2}\biggl({\lambda_7^2\over M_U^2}
+{\lambda_8^2\over M_D^2}
+{\lambda_9^2+\lambda_{10}^2\over M^2}\biggr)\biggr]\nonumber \\ &=&- c_{gg}.
\label{yukdef}
\end{eqnarray}
The top and bottom quark couplings to $ggh$ and $gghh$ are not included in $c_g$ and $c_{gg}$,
but can be calculated at one-loop using the effective interactions of Eq. \ref{effective}. 
The effective Lagrangian depends on only $3$ new parameters: $c_g, \delta_t$, and $ \delta _b$,
along with the physical masses, $m_t=M_{T_1}$ and $ m_b=M_{B_1}$, and $v$.  It is important
to note, that within the context of this model, the coefficients of the effective Lagrangian cannot
all be independently varied.  This feature can also arise in composite Higgs models~\cite{Grober:2010yv}.
 
The non-Standard Model couplings to the $W$ and $Z$ are given to ${\cal O}(1/M^2_X)$ by,
\begin{eqnarray}
&\delta g_L=-{v^2\over 4}
\biggl(
{\lambda_7^2\over M_U^2}+{\lambda_8^2\over M_D^2}\biggr)
\qquad\qquad
&\delta g_R= {v^2\lambda_9 \lambda_{10} \over 2  M^2}
\nonumber \\
&\delta Z_L^t= -{v^2 \lambda_7^2 \over 4 M_U^2}
\qquad\qquad
&\delta Z_R^t= {v^2 \lambda_9^2 \over 4 M^2}
\nonumber \\
&\delta Z_L^b= {v^2 \lambda_8^2 \over 4 M_D^2}
\qquad\qquad
&\delta Z_R^b=- {v^2 \lambda_{10}^2 \over 4 M^2}
\label{coupsw}
\end{eqnarray}

\section{Understanding the Full Theory} 
\label{smallmat}
\subsection{Hierarchy 1}
\label{hier1sec}
 In order to understand some general features of the mass matrices we consider a hierarchy
 where the mixing angles are small,
\begin{eqnarray}
\theta\sim \frac{\lambda_i v}{M_4}
\sim \frac{\lambda_i v}{M_5}
\sim \frac{M_4}{M}
\sim \frac{M_5}{M}
\quad{\rm and}\quad
 \theta^2\sim \frac{\lambda_i v}{M}\, .
 \label{angscale}
\end{eqnarray}
This maintains the hierarchy, $\lambda_i v \ll  M_4,M_5 \ll  M,M_U,M_D$, keeping the
off-diagonal elements of the mass matrices small. 
In this limit the matrices which diagonalize the top quark mass matrix can be written as 
\begin{eqnarray}
V_L^t &=& \begin{pmatrix} 1-\frac{1}{2}\theta_L^{D2} & 
- \theta_L^D &
 - \theta_L^{S2} \\ 
 \theta_L^D & 1-\frac{1}{2}\theta_L^{D2} 
 & \theta_L^{H2} \\ 
 \theta_L^{S2} & - \theta_L^{H2} & 1\end{pmatrix}\nonumber\\
 V_R^t &=& \begin{pmatrix}
 1 -\theta_R^{S2} & 
 -\theta_R^{D2} & 
 -\theta_R^S \\
  \theta_R^{D2} & 1 & 
  -\theta_R^{H2} \\
   \theta_R^S & \theta_R^{H2} &
   1-\frac{\theta_R^{S2}}{2}
    \end{pmatrix}\, ,
\label{msmall}
\end{eqnarray}
where the matrices of Eq. \ref{msmall} are unitary to ${\cal{O}}(\theta^3)$.\footnote{Note that the hierarchy determines the leading behavior of the $\theta$ expansion of the mixing matrices.  Higher orders of this expansion are determined by unitarity.}  
The angles $\theta^D$ ($\theta^S$) can be thought of as the doublet (singlet) vector fermion-mixing with the Standard Model-like top quark,
and $\theta^H$ as the doublet-singlet vector fermion mixing.  All angles are assumed to scale as Eq. \ref{angscale}.

In the small angle limit of Eq. \ref{msmall}, we can 
then solve for the parameters of the Lagrangian:
\begin{eqnarray}\mathcal{M}^{(t)}(0)&=&
\begin{pmatrix} \lambda_t\frac{v}{\sqrt{2}} & M_4 &\lambda_7\frac{v}{\sqrt{2}}\\ \lambda_9\frac{v}{\sqrt{2}} & M & \lambda_1\frac{v}{\sqrt{2}} \\ M_5 & \lambda_3 \frac{v}{\sqrt{2}} & M_U \end{pmatrix}\nonumber \\
&=& \begin{pmatrix} M_{T_1}\left(1-\frac{{\theta^D_L}^2}{2}-\frac{{\theta^S_R}^2}{2}\right) & M_{T_2}\theta_L^D-M_{T_1}{\theta_R^D}^2  &M_{T_3}{\theta_L^S}^2 -M_{T_1}\theta_R^S   \\
 M_{T_2}{\theta_R^D}^2-M_{T_1}\theta_L^D  & M_{T_2}\left( 1-\frac{{\theta^D_L}^2}{2}\right) &-M_{T_3}{\theta_L^H}^2 - M_{T_2}{\theta_R^H}^2+ M_{T_1}\theta_L^D\theta_R^S \\ 
M_{T_3}\theta_R^S-M_{T_1}{\theta_L^S}^2  & M_{T_2}{\theta_L^H}^2 + M_{T_3}{\theta_R^H}^2 & M_{T_3}\left(1-\frac{{\theta^S_R}^2}{2}\right)\end{pmatrix}\nonumber
\label{gaugeSol}
\end{eqnarray}
As can be seen, according to the $\theta$ scaling behaviour, this obeys the structure that we want ($\lambda_i v\ll \lambda_{4,5} \ll M$).  
In the fermion mass-eigenstate basis,
\begin{equation}
L\sim -\sum_{i,j=1,2,3}{\overline \psi}^t_i {\cal Y}_{ij}^t \psi^t_j h\, ,
\label{yf}
\end{equation}
the small angle approximation to the charge ${2\over 3}$ Yukawa interactions is,
\begin{eqnarray}
v\times {\cal Y}^t &=& \begin{pmatrix} M_{T_1} & M_{T_1} {\theta_R^D}^2 & M_{T_3} {\theta_L^S}^2 \\  M_{T_2} {\theta_R^D}^2 & 0 & -M_{T_3}{\theta_L^H}^2 - M_{T_2} {\theta_R^H}^2 \\ M_{T_1} {\theta^S_L}^2 & M_{T_2} {\theta_L^H}^2+ M_{T_3} {\theta_R^H}^2 & 0 \end{pmatrix}\, .
\end{eqnarray}
The mass matrix in the $b$ quark sector can be parameterized in an identical fashion to the above discussion.

The $W$ interactions defined in Eqs.~\ref{wcoups} and \ref{wmatdef}, in the small angle approximation of Eq. \ref{msmall},
are
\begin{eqnarray}
U_L&=&\begin{pmatrix}1-\frac{1}{2}\left(\theta_L^{Db}-\theta_L^{Dt}\right)^2&\theta_L^{Db}-\theta_L^{Dt}&{\theta_L^{Sb}}^2 \\ \theta_L^{Dt}-\theta_L^{Db} & 1-\frac{1}{2}\left(\theta_L^{Db}-\theta_L^{Dt}\right)^2 & -{\theta_L^{Hb}}^2 \\ {\theta_L^{St}}^2 & -{\theta_L^{Ht}}^2 & 0 \end{pmatrix}\\
 U_R &=& \begin{pmatrix}0 & -{\theta_R^{Dt}}^2 & 0 \\ -{\theta_R^{Db}}^2 & 1 & {\theta_R^{Hb}}^2 \\ 0 & {\theta_R^{Ht}}^2 & 0\end{pmatrix},
\end{eqnarray}
where we have added the superscripts $b,t$ to indicate mixing in the bottom and top sectors, respectively.  
The $Z$-fermion interactions defined in Eqs.~\ref{zcoup} and \ref{zcoupdef}, in the small angle approximation are
\begin{eqnarray}
 X_L^t&=&\begin{pmatrix} 1 & 0 & {\theta_L^{St}}^2 \\ 0 & 1 & -{\theta_L^{Ht}}^2 \\  {\theta_L^{St}}^2 & - {\theta_L^{Ht}}^2 & 0\end{pmatrix}\\
 X^t_R &=&\begin{pmatrix}0 & -{\theta_R^{Dt}}^2 & 0\label{Zmix1} \\ -{\theta_R^{Dt}}^2 & 1 & {\theta_R^{Ht}}^2 \\ 0 & {\theta_R^{Ht}}^2 & 0\end{pmatrix}\, .\nonumber
\end{eqnarray}
The results for the bottom sector can be found by the replacement $t\rightarrow b$.

Comparing to the EFT of Eq.  \ref{effective} in the small
angle approximation described above,
\begin{eqnarray}
Y_t&=&\frac{M_{T_1}}{v}\nonumber\\
Y_b&=&\frac{M_{B_1}}{v}\nonumber\\
c_{2h}^{(t)}&=&c_{2h}^{(b)}=0\nonumber\\
c_g&=&-c_{gg}=0\, .
\label{EFThier1}
\end{eqnarray}
In the EFT, this hierarchy reduces to the Standard Model and so does not produce large deviations in Higgs production rates.

\subsection{Hierarchy 2}
\label{hier2sec}
Hierarchy $1$ appears to give small $\lambda_7,\lambda_9,\lambda_3,\lambda_8,\lambda_{10},\lambda_{11}$, which are the parameters that give  deviations from the Standard Model.  
We now  describe a different hierarchy with  $M_{4,5}\ll \lambda_i  v\ll M$:
\begin{eqnarray}
\theta\sim \frac{M_{4,5}}{\lambda_iv}\sim \frac{\lambda_i v}{M}\quad{\rm and}\quad \theta^2\sim \frac{M_{4,5}}{M}\, .
\end{eqnarray}
The diagonalization matrices can be parameterized  in both the $t$ sector as\footnote{We
omit the superscript $t$ and $b$ on the mixing angles where it is obvious.},
\begin{eqnarray}
V_L^t &=& \begin{pmatrix} 1-\frac{1}{2}{\theta_L^S}^2 & - {\theta_L^D}^2 & -\theta_L^S \\ {\theta_L^D}^2+\theta_L^H\theta_L^S & 1-\frac{1}{2}{\theta_L^H}^2 & \theta_L^H \\ \theta_L^S & -\theta_L^H & 1-\frac{1}{2} \left({\theta_L^S}^2+{\theta_L^H}^2\right)\end{pmatrix}\nonumber \\
 V_R^t &=& \begin{pmatrix} 1-\frac{1}{2}{\theta_R^D}^2 & -\theta_R^D & -{\theta_R^S}^2 \\ \theta_R^D & 1-\frac{1}{2}\left({\theta_R^D}^2+{\theta_R^H}^2\right) & -\theta_R^H \\ \theta_R^D \theta_R^H+{\theta_R^S}^2 & \theta_R^H & 1-\frac{1}{2}{\theta_R^H}^2 \end{pmatrix}\, .
 \label{hier2mat}
\end{eqnarray}
The parameters of the original top mass matrix, $M^{(t)}(0)$ from Eq.~\ref{massdef}, can be solved for to ${\cal O}(\theta^2)$,
\begin{eqnarray}
\lambda_t\frac{v}{\sqrt{2}} &=&M_{T_1}\left(1-\frac{{\theta^S_L}^2}{2}-\frac{{\theta^D_R}^2}{2}\right) \nonumber\\
M_4 &=&  M_{T_2}\left({\theta_L^D}^2+\theta_L^S\theta_L^H\right)+M_{T_3}\theta_L^S\theta_R^H-M_{T_1}\theta_R^D\nonumber\\
\lambda_7\frac{v}{\sqrt{2}} &=& M_{T_3}\theta_L^S-M_{T_1}{\theta_R^S}^2 \nonumber\\
\lambda_9\frac{v}{\sqrt{2}} &=& M_{T_2}\theta_R^D-M_{T_1}{\theta_L^D}^2\nonumber \\
 M &=&M_{T_2}\left[1 -\frac{1}{2}\left({\theta_L^H}^2+{\theta_R^D}^2+{\theta_R^H}^2\right)\right]-M_{T_3}\theta_L^H\theta_R^H  \nonumber\\
\lambda_1\frac{v}{\sqrt{2}} &=&-M_{T_3}\theta_L^H-M_{T_2}\theta_R^H \nonumber\\
M_5 &=& M_{T_3}(\theta_R^D\theta_R^H+{\theta_R^S}^2)+M_{T_2}\theta_R^D\theta_L^H -M_{T_1}\theta_L^S \nonumber \\
\lambda_3 \frac{v}{\sqrt{2}} &=&M_{T_3}\theta_R^H+M_{T_2}\theta_L^H+M_{T_1}\theta_L^S\theta_R^D \nonumber \\
M_U &=& M_{T_3}\left[1-\frac{1}{2} \left({\theta_L^S}^2+{\theta_L^H}^2+{\theta_R^H}^2\right)\right]-M_{T_2}\theta_L^H\theta_R^H
\end{eqnarray}
Finally, the Higgs couplings to the charge ${2\over 3}$ fermions can be written as 
in Eq. \ref{yf},
\begin{eqnarray} 
&&
v\times {\cal Y }^t=
\nonumber \\
&&\hskip -1.1in
\begin{pmatrix}
M_{T_1}\left(1-{\theta_L^S}^2-{\theta_R^D}^2\right)
& M_{T_1} \theta_R^D-M_{T_2} \theta_L^H\theta_L^S-2 M_{T_3} \theta_L^S\theta_R^H
&M_{T_1} \theta_R^D\theta_R^H +M_{T_3} \theta_L^S\nonumber \\
M_{T_1} \theta_L^H\theta_L^S+M_{T_2} \theta_R^D
&M_{T_2} \left({\theta_L^H}^2+{\theta_R^D}^2+{\theta_R^H}^2\right)+2 M_{T_3} \theta_L^H\theta_R^H
&-M_{T_3} \theta_L^H - M_{T_2}\theta_R^H\nonumber \\
M_{T_1} \theta_L^S - 2 M_{T_2} \theta_L^H \theta_R^D-M_{T_3} \theta_R^D\theta_R^H
&2 M_{T_1} \theta_L^S \theta_R^D + M_{T_2} \theta_L^H + M_{T_3} \theta_R^H
&M_{T_3}\left({\theta_L^H}^2+{\theta_L^S}^2+{\theta_R^H}^2\right)+ 2 M_{T_2} \theta_L^H \theta_R^H
\nonumber 
\end{pmatrix}
\\ 
\end{eqnarray}
Again, the $b$ sector mass matrix and mixing can be parameterized in a similar fashion to the above.

Comparing to the EFT in Eq. \ref{effective} (counting $M_{T_1}/M_{T_{2,3}}\sim M_{B_1}/M_{B_{2,3}}\sim \theta$), this hierarchy
yields small deviations from the Standard Model,
\begin{eqnarray}
Y_t&=&\frac{M_{T_1}}{v}\left(1-{\theta_R^{Dt}}^2-{\theta_L^{St}}^2\right)\nonumber\\
Y_b&=&\frac{M_{B_1}}{v}\left(1-{\theta_R^{Db}}^2-{\theta_L^{Sb}}^2\right)\nonumber\\
c_{2h}^{(t)}&=&\frac{3M_{T_1}}{2v^2}\left({\theta_R^{Dt}}^2+{\theta_L^{St}}^2\right)\nonumber\\
c_{2h}^{(b)}&=&\frac{3M_{B_1}}{2v^2}\left({\theta_R^{Db}}^2+{\theta_L^{Sb}}^2\right)\nonumber\\
c_g=-c_{gg}&=& (2{\theta_L^{Ht}}^2+{\theta_L^{St}}^2)+(2{\theta_R^{Ht}}^2+{\theta_R^{Dt}}^2)+2\frac{M^2_{T_2}+M_{T_3}^2}{M_{T_2}M_{T_3}}\theta_L^{Ht}\theta_R^{Ht}\nonumber\\
&&+(2{\theta_L^{Hb}}^2+{\theta_L^{Sb}}^2)+(2{\theta_R^{Hb}}^2+{\theta_R^{Db}}^2)+2\frac{M^2_{B_2}+M_{B_3}^2}{M_{B_2}M_{B_3}}\theta_L^{Hb}\theta_R^{Hb} \, .
\label{hier2EFT}
\end{eqnarray}
Again, the superscripts $t,b$ indicate mixing angles in the top and bottom sectors, respectively.  
Now we want to match onto the LET, i.e, integrate out the top quark ($T_1$), along with the heavier fermions 
$T_2,T_3,B_2$, and  $B_3$. 
The effective Higgs-gluon interactions are
\begin{eqnarray}
\mathcal{L}_{\rm LET} = \frac{\alpha_s}{12\pi}\left[(1+c_g^{\rm LET})\frac{h}{v}-\frac{1+c_{gg}^{\rm LET}}{2}\frac{h^2}{v^2}\right]G^{A,\mu\nu}G^A_{\mu\nu}\, .
\end{eqnarray}
To obtain $c^{LET}$ we use the full LET
 for  the top quark sector in Eqs.~\ref{EFTfull} and ~\ref{ggHHFull}, then add in the effect of integrating out the heavy bottom quark partners,   
 that is, the heavy down-type quark contributions to $c_g$ and $c_{gg}$ in Eq.~\ref{yukdef}.
 To ${\cal O}(\theta^2)$, this yields,
\begin{eqnarray}
c_g^{LET}&=&-c_{gg}^{LET}\label{cgHier2}\\
&=&2\left({\theta_L^{Hb}}^2+{\theta_L^{Ht}}^2+{\theta_R^{Hb}}^2+{\theta_R^{Ht}}^2\right)+{\theta_L^{Sb}}^2+{\theta_R^{Db}}^2+2\frac{M^2_{T_2}+M_{T_3}^2}{M_{T_2}M_{T_3}}\theta_L^{Ht}\theta_R^{Ht}+2\frac{M^2_{B_2}+M_{B_3}^2}{M_{B_2}M_{B_3}}\theta_L^{Hb}\theta_R^{Hb}\nonumber.
\end{eqnarray}
For degenerate heavy fermions, $c_g$ is positive definite and so the contribution to double Higgs
production from $c_{gg}$ always decreases the rate. Additionally, to increase the double Higgs contribution from $c_{gg}$, $\theta^H_L$ and $\theta^H_R$ should have opposite signs. 

The mixing matrices for  the $W$ interactions are (Eqs.~\ref{wcoups} and \ref{wmatdef}).
\begin{eqnarray}
U_L &=& \begin{pmatrix} 1-\frac{1}{2}\left({\theta_L^{Sb}}^2+{\theta_L^{St}}^2\right) & {\theta_L^{Db}}^2-{\theta_L^{Dt}}^2+\theta_L^{Hb}\theta_L^{Sb} & \theta_L^{Sb}\\
{\theta_L^{Dt}}^2-{\theta_L^{Db}}^2+\theta_L^{Ht}\theta_L^{St} &1-\frac{1}{2}\left({\theta_L^{Hb}}^2+{\theta_L^{Ht}}^2\right) & -\theta_L^{Hb}\\
\theta_L^{St} & -\theta_L^{Ht} & \theta_L^{Hb}\theta_L^{Ht}+\theta_L^{Sb}\theta_L^{St}\end{pmatrix}\label{Wmix}\\
U_R&=& \begin{pmatrix} \theta_R^{Db} \theta_R^{Dt} & -\theta_R^{Dt} & -\theta_R^{Hb} \theta_R^{Dt}\\ -\theta_R^{Db} & 1-\frac{1}{2}\left({\theta_R^{Db}}^2+{\theta_R^{Hb}}^2+{\theta_R^{Dt}}^2+{\theta_R^{Ht}}^2\right) & \theta_R^{Hb} \\
-\theta_R^{Db}\theta_R^{Ht} & \theta_R^{Ht} & \theta_R^{Hb}\theta_R^{Ht} \end{pmatrix}\nonumber
\end{eqnarray}
The mixing matrices for $Z$ interactions (Eqs.~\ref{zcoup} and \ref{zcoupdef}) in the charge ${2\over 3}$ sector are,
\begin{eqnarray}
X_L^t &=& \begin{pmatrix} 1-{\theta_L^{St}}^2 & \theta_L^{Ht}\theta_L^{St} & \theta_L^{St}\\
\theta_L^{Ht}\theta_L^{St} &1-{\theta_L^{Ht}}^2 & -\theta_L^{Ht}\\
\theta_L^{St} & -\theta_L^{Ht} & {\theta_L^{Ht}}^2+{\theta_L^{St}}^2\end{pmatrix}\label{Zmix}\\
X_R^t&=& \begin{pmatrix} {\theta_R^{Dt}}^2 & -\theta_R^{Dt} & -{\theta_R^{Ht}}^2\\ -\theta_R^{Dt} & 1-{\theta_R^{Dt}}^2-{\theta_R^{Ht}}^2 & \theta_R^{Ht} \\
-\theta_R^{Dt}\theta_R^{Ht} & \theta_R^{Ht} & {\theta_R^{Ht}}^2 \end{pmatrix}\nonumber
\label{zzt}
\end{eqnarray}
 The $Z$ couplings in the bottom sector are found from Eq. \ref{Zmix}  with the replacement  $t\rightarrow b$.

%%%%%%%%%%%%%%%%%%%%%%%%%%%%%%%%%%%%%%%%%%%%%%%%%%%%%%%%%%%%%%%%%%%%%%%%%%%

\section{Limits from Precision Measurements}
\label{secresults}
New heavy quarks which couple to the Standard Model gauge bosons are 
restricted by the oblique parameters~\cite{Peskin:1991sw}.  In addition, the couplings of charge $-{1\over 3}$ quarks 
are significantly limited by the measurements of  $Z\rightarrow b {\overline b}$.  
These  limits typically require small mixing parameters.  
. 

General formulas for the contributions of the fermion sector to $\Delta S$ and $\Delta T$ are
given in Appendix A.  It is useful to consider several special cases here.
For the case with only a top partner singlet ($T_3$) with a mass $M_{T_3}\gg M_{T_1}$, the only non-zero entries of 
the left-handed mixing matrices are,
\begin{eqnarray}
V_{L,11}^t&=& V_{L,33}^t=c_L\nonumber \\
V_{L,31}^t&=& -V_{L,13}^t=-s_L\nonumber \\
V_{L,11}^b&=&1\, ,
\end{eqnarray}
while $V_R^{t,b}$ can be set to the unit matrix, $c_L\equiv \cos\theta_L$, $s_L\equiv\sin\theta_L$, and
\begin{eqnarray}
\tan (2\theta_L)={\sqrt{2}v\over M_U}\biggl({\lambda_7\over 1-(\lambda_7^2+\lambda_t^2){v^2\over 2 M_U^2}}\biggr)
\end{eqnarray}
The result for large top partner masses is (after subtracting the Standard Model top and bottom
contributions),
\begin{eqnarray}
\biggl[\Delta T\biggr]_{top~singlet}&=& {N_c\over 16\pi s_W^2 M_W^2}s_L^2\biggl(
-(1+c_L^2)M_{T_1}^2-2c_L^2 M_{T_1}^2\ln\biggl({M_{T_1}^2\over M_{T_3}^2}\biggr)+s_L^2 M_{T_3}^2\biggr)
\nonumber \\
\biggl[\Delta S\biggr]_{top~singlet}&=& -{N_c\over 18\pi }s_L^2\biggl(
5c_L^2
+(1-3c_L^2)\ln\biggl({M_{T_1}^2\over M_{T_3}^2}\biggr)\biggr)\, ,
\label{tsing}
\end{eqnarray}
where $N_C=3$, 
in agreement with Ref. \cite{Dawson:2012mk},
which found that  fits to the oblique
parameters  require ${s_L\lsim 0.16}$ for $M_{T_3}\sim 1$~TeV at $95\%$ confidence level.  For fixed values
of the Yukawa couplings, $\lambda_i$, the mixing angle scales for large $M_{T_3}$  as,
\begin{equation}
s_L\sim {v\lambda_i\over M_{T_3}}
\end{equation}
and the contributions to the oblique parameters from the top partner decouple,
\begin{equation}
\biggl[\Delta T\biggr]_{top~singlet}\sim \biggl[\Delta S\biggr]_{top~singlet}
\sim {\lambda_i^2 v^2\over M_{T_3}^2}\, .
\end{equation}

The limit on the angle $s_L$ in the above example arises because of the mixing with the Standard Model
top quark.  Ref. \cite{Dawson:2012mk}  contains
 an example where there is  a heavy vector-like $SU(2)_L$ doublet, $Q$,   along
with vector-like charge ${2\over 3}$ and $-{1\over 3}$ quarks, $U$ and $D$, which are not allowed to mix
with the Standard Model fermions.  This corresponds to $M_4=M_5=M_6=\lambda_7=\lambda_8=\lambda_9
=\lambda_{10}=0$ in Eq. \ref{modeldef}.  In this case, limits from the oblique parameters require that the heavy 
fermions be approximately degenerate, $M_{T_2}\simeq M_{T_3}\simeq M_{B_2}\simeq M_{B_3}$, while one combination of
mixing angles is unconstrained.   

Limits can be also  obtained from $Z$ decays to $b {\overline b}$ by  
comparing the experimental result\cite{Beringer:1900zz}
for $R_b$ with the recent Standard Model calculation\cite{Freitas:2014hra}, 
\begin{eqnarray}
R_b&\equiv &
{\Gamma(Z\rightarrow b {\overline b})
\over \Gamma(Z\rightarrow b {\overline b})}
\nonumber \\
R_b^{exp}&=& 0.21629\pm 0.00066\nonumber \\
R_b^{SM}&=& 0.2154940.
\end{eqnarray}
$R_b$ can be related to the anomalous couplings of the $b$ quark to the $Z$ given
in Eq. \ref{effective}
\begin{equation}
{R_b^{exp}\over R_b^{SM}}=1-3.57 \delta g_L^b+0.65 g_R^b\, .
\label{rbdef}
\end{equation}
From Eqs.  \ref{coupsw} and \ref{rbdef}, we extract
 the $95\%$ confidence level bound,
\begin{equation}
\biggl({M\over  \lambda_8}\biggr)^2
\biggl({1\over 
1+0.224(1-{\lambda_{10}M\over \lambda_8M_D})^2}\biggr)\gsim
( 2~{\rm TeV})^2\, .
\end{equation}

The following discussion focuses on Hierarchy 2 of Section \ref{smallmat}, although it can be shown that the conclusions are quite generic. We start by
counting  the degrees of freedom.  Naively, there are  $6$ masses,
\begin{eqnarray}
M_{T_1},\,M_{T_2},\,M_{T_3},\,M_{B_1},\,M_{B_2},\,M_{B_3}
\end{eqnarray}
and 12 angles,
\begin{eqnarray}
\theta_{L,R}^{St},\,\theta_{L,R}^{Dt},\,\theta_{L,R}^{Ht},\,\theta_{L,R}^{Sb},\,\theta_{L,R}^{Db},\,\theta_{L,R}^{Hb}
\, .
\end{eqnarray}
However, $M_4$ and $M$ are the same in the top and bottom sectors, leaving a 
total of $16$ independent
parameters.
Considering Eqs.~\ref{Zmix1} and \ref{Zmix}, we see that if we forbid mixing between particles with different quantum numbers then flavor changing neutral currents
involving the Z are eliminated.  That is, $\theta_L^{St}$ mixes a component of the Standard Model $SU(2)_L$ doublet with an $SU(2)_L$ singlet, and $\theta_R^{Dt}$ mixes a Standard Model $SU(2)_L$ singlet with a component of a vector fermion doublet.  We set these angles to zero to avoid restrictions from deviations in the $3^{rd}$ generation quark neutral current couplings, in particular $Z\rightarrow b \overline{b}$:
\begin{eqnarray}
\theta_L^{St}=\theta_R^{Dt}=\theta_L^{Sb}=\theta_R^{Db}=0. \label{EWPOconstraints}
\end{eqnarray}
The angles $\theta^{Ht}_{L,R}$ and $\theta^{Hb}_{L,R}$ are left nonzero, since from Eq.~\ref{cgHier2} we see that these are intimately tied to deviations from Standard Model Higgs production rates. 
The $Z$ couplings to the top quark and heavy up-type vector quarks are then
\begin{eqnarray}
X_L^t &=& \begin{pmatrix} 1 & 0 & 0\\
0 &1-{\theta_L^{Ht}}^2 & -\theta_L^{Ht}\\
0 & -\theta_L^{Ht} & {\theta_L^{Ht}}^2+{\theta_L^{St}}^2\end{pmatrix}\label{ZDiagmix}\\
X_R^t&=& \begin{pmatrix} 0 & 0 & 0\\ 0 & 1-{\theta_R^{Ht}}^2 & \theta_R^{Ht} \\
0 & \theta_R^{Ht} & {\theta_R^{Ht}}^2 \end{pmatrix}\nonumber\, ,
\end{eqnarray}
and the $t$ and $b$ quarks have Standard Model-like neutral current couplings.

The $W$-mixing matrices in Hierarchy 2 are,
\begin{eqnarray}
U_L &=& \begin{pmatrix} 1 & {\theta_L^{Db}}^2-{\theta_L^{Dt}}^2 & 0\\
{\theta_L^{Dt}}^2-{\theta_L^{Db}}^2 &1-\frac{1}{2}\left({\theta_L^{Hb}}^2+{\theta_L^{Ht}}^2\right) & -\theta_L^{Hb}\\
 0& -\theta_L^{Ht} & \theta_L^{Hb}\theta_L^{Ht}\end{pmatrix}\label{Wmix1}\\
U_R&=& \begin{pmatrix} 0 & 0 & 0\\ 0 & 1-\frac{1}{2}\left({\theta_R^{Hb}}^2+{\theta_R^{Ht}}^2\right) & \theta_R^{Hb} \\
0 & \theta_R^{Ht} & \theta_R^{Hb}\theta_R^{Ht} \end{pmatrix}\nonumber\, .
\end{eqnarray}
$U_R$ only depends on $\theta^{Ht}_{L,R}$ and $\theta^{Hb}_{L,R}$, the mixing angles between the heavy vector fermions, while $U_L$ still depends on the mixing between the heavy states with the Standard Model.   Forcing the heavy-light mixing to be isospin conserving, $\theta_L^{Db}=\theta_L^{Dt}$, $U_L$ becomes
\begin{eqnarray}
U_L &=& \begin{pmatrix} 1 & 0 & 0\\
0 &1-\frac{1}{2}\left({\theta_L^{Hb}}^2+{\theta_L^{Ht}}^2\right) & -\theta_L^{Hb}\\
 0& -\theta_L^{Ht} & \theta_L^{Hb}\theta_L^{Ht}\end{pmatrix}\label{Wmix2}
\end{eqnarray}
and there are no gauge boson currents mixing the Standard Model top and bottom quarks with the new vector fermions.  

To summarize, taking into consideration electroweak precision observables, it is reasonable to impose the constraints:
\begin{eqnarray}
\theta_L^{St}=\theta_R^{Dt}=\theta_L^{Sb}=\theta_R^{Db}=0,\quad\theta_L^{Db}=\theta_L^{Dt}.
\end{eqnarray}
Under this assumption, the non-zero mixing angles are,
\begin{eqnarray}
\theta_{R}^{St},\,\theta_{L}^{Dt},\,\theta_{L,R}^{Ht},\,\theta_{R}^{Sb},\,\theta_{L,R}^{Hb},
\end{eqnarray}
There are $2$  constraints from $M_4$ and $M$,
\begin{eqnarray}
M_4&=&  M_{T_2}{\theta_L^{Dt}}^2=M_{B_2}{\theta_L^{Db}}^2\nonumber\\
 M &=&M_{T_2}\left(1-\frac{1}{2} {\theta_L^{Ht}}^2-\frac{1}{2}{\theta_R^{Ht}}^2\right)-M_{T_3}\theta_L^{Ht}\theta_R^{Ht}  \nonumber\\
 &=&M_{B_2}\left(1-\frac{1}{2} {\theta_L^{Hb}}^2-\frac{1}{2}{\theta_R^{Hb}}^2\right)-M_{B_3}\theta_L^{Hb}\theta_R^{Hb}.
 \label{rest}
\end{eqnarray}
So, $\theta_L^{Dt}=\theta_L^{Db}$ is only consistent  if $M_{T_2}=M_{B_2}$, which fully eliminates isospin violation in the mixing between the new heavy states and the $3^{rd}$ generation quarks.  To make things simpler, we can also assume $M_{T_3}=M_{B_3}$, and then Eq. \ref{rest} is satisfied when  $\theta_{L}^{Ht}=\theta_L^{Hb}$ and $\theta_R^{Ht}=\theta_R^{Hb}$.  (There are other possible solutions not requiring $M_{T_3}=M_{B_3}$, but for simplicity we focus on this limit.)

Now we only have a few remaining degrees of freedom:  4 masses (2 of which are known)
\begin{eqnarray}
M_{T_1},\,M_{B_1},\,M_{T_2}=M_{B_2},\,M_{T_3}=M_{B_3}
\end{eqnarray}
and five angles,
\begin{eqnarray}
\theta_R^{St},\,\theta_R^{Sb},\,\theta_L^{Dt}=\theta_L^{Db},\,\theta_L^{Ht}=\theta_L^{Hb},\,\theta_R^{Ht}=\theta_R^{Hb}.
\end{eqnarray}
 At lowest order these angles are unconstrained by $Z\rightarrow b\overline{b}$ and the oblique parameters only constrain the mixing among the heavy quarks.  These constraints can be found in Ref.~\cite{Dawson:2012mk}.  Although this result can be shown generically without assuming that $\theta^{D}_L$ and $\theta^S_R$ are small, these angles will manifest themselves in the CKM matrix when considering mixing among the first three generations~\cite{Ellis:2014dza}.  We therefore continue with the small angle approximation.

\section{Results for Higgs Production}
\label{higgsres}
In this section, we compare the accuracy of the low energy theorem (LET) with the effective Lagrangian
obtained by including the top and bottom quark mass effects (EFT), Eq.~\ref{effective},
 as well as with predictions obtained using the
full theory.  
We have two goals: the first is to understand the numerical limitations of the approximations to the full theory.
Our second goal is to search for  a regime where single Higgs production from gluon fusion
occurs at approximately the Standard Model rate, while double Higgs production
is significantly altered.  Again, we focus on Hierarchy 2 of Section~\ref{hier2sec}, since Hierarchy 1 (Section~\ref{hier1sec}) does not lead to significant deviations from the Standard Model (Eq.~\ref{EFThier1}).

We normalize the predictions to the Standard Model rates,
\begin{eqnarray}
R_h&\equiv&{\sigma(gg\rightarrow h)\over \sigma(gg\rightarrow h)_{SM}}
\nonumber \\
R_{hh}&=&{\sigma(gg\rightarrow hh)\over \sigma(gg\rightarrow hh)_{SM}}.
\label{rdefs}
\end{eqnarray}
To ${\cal O}(\delta_{LET})$, the low energy theorems of Eqs. \ref{letggh} and \ref{letgghh}, including only the up-type quarks, predict,
\begin{eqnarray}
R_h&\sim& 1+2\delta_{LET}
\nonumber\\
R_{hh}&\sim&1+2\delta_{LET}-{4\delta_{LET}\over F_0^{SM}(M_{T_1}\rightarrow\infty)}\, ,
\label{rhlet}
\end{eqnarray}
and 
\begin{equation}
F_0^{SM}(M_{T_1}\rightarrow\infty)\equiv 1-{3M_h^2\over s-M_h^2}\, ,
\label{gghhlet}
\end{equation}
where $\delta_{LET}=2\lambda_3 v^2(\lambda_1\lambda_t-\lambda_7\lambda_9)/X$
 is given in Eq. \ref{letggh} and $F_0$ is defined in Eqs.~\ref{eq:amp}, \ref{eftlim}, \ref{eftgghh}.  In the effective field theory language of Eq. \ref{effective}, $\delta_{LET}=c_g$.
The presence of the $\lambda_3$ coupling does indeed allow single Higgs production to 
differ from the Standard Model prediction.  However, once $R_h$ is measured to be
approximately $1$, the deviations of $R_{hh}$ from $1$ are restricted to be small.   Thus
in order for the double Higgs rate to be different from the Standard Model prediction, we
need a region of parameter space where the low energy theorem is not valid.

The rate for single Higgs production in the effective theory including all top  and
bottom quark mass effects (EFT), but
integrating out the heavy vector-like fermions to ${\cal O}\biggl({1\over M^2_X}\biggr)$
and assuming $\delta_b,\delta_t$ and $c_g$ are small, is given by,
\begin{eqnarray}
R_h &\rightarrow
&{\mid (1+\delta_t)F_{1/2}(\tau_{T_1})+(1+\delta_b)F_{1/2}(\tau_{B_1})+c_{g}F_{1/2}^\infty\mid^2
\over 
\mid F_{1/2}(\tau_{T_1})+F_{1/2}(\tau_{B_1})\mid^2}
\nonumber \\
&\sim&1+
2 \biggl[
{ \delta_t \mid F_{1/2}(\tau_{T_1})\mid^2
+\delta_b\mid F_{1/2}(\tau_{B_1})\mid^2 +
(\delta_t+\delta_b)
Re\biggl( F_{1/2}(\tau_{T_1})F^*_{1/2}(\tau_{B_1})\biggr)
\over 
\mid F_{1/2}(\tau_{T_1})+F_{1/2}(\tau_{B_1})\mid^2}\biggr]
\nonumber \\
&&
+2\biggl[{c_{g}F_{1/2}^\infty Re\biggl(F_{1/2}(\tau_{T_1})+F_{1/2}(\tau_{B_1})\biggr)
\over 
\mid F_{1/2}(\tau_{T_1})+F_{1/2}(\tau_{B_1})\mid^2}\biggr]\, ,
\end{eqnarray}
where $\tau_i\equiv 4M_i^2/M_h^2$,  
\begin{eqnarray}
F_{1/2}(\tau)&=&-2\tau[1+(1-\tau)f(\tau)]\nonumber \\
f(\tau)&=& \left\{ \begin{matrix*}[l]\biggl[
\sin^{-1}\biggl({1\over \sqrt{\tau}}\biggr)\biggr]^2\quad&\hbox{if~} \tau\ge 1 \\
 -{1\over 4}
\biggl[
\ln\biggl(
{1+\sqrt{1-\tau}\over 1-\sqrt{1-\tau}}\biggr)-i\pi\biggr]^2\quad &{\hbox {if}}~\tau < 1\,\end{matrix*}\right. ,
\end{eqnarray}
and $F_{1/2}^{\infty}=-{4\over 3}$ 
in the $M_{T_1}\rightarrow \infty$ limit of 
$F_{1/2}(\tau_{T_1})$.
Neglecting the $b$ contribution and noting that $F_{1/2}(\tau_{T_1})$ is well approximated
by $F_{1/2}^\infty$, 
\begin{equation}
R_h\sim 1+2(\delta_t+c_g)\, .
\label{singans}
\end{equation}
The $c_g$ contribution is in agreement with the LET result of Eq. \ref{rhlet}.

\begin{figure}[tb]
%\vskip -1in.
\begin{center}
%\hskip -.5in
\includegraphics[scale=1.2,clip]{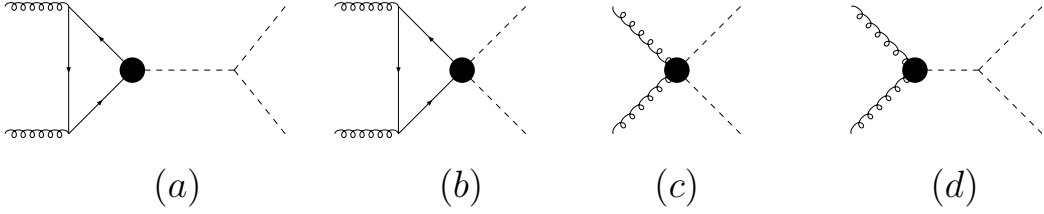}
%\vskip -7. in
\caption{Non-box contributions to the spin-0 component of $gg\rightarrow hh$.  The dark circles represent the non-Standard Model contributions, 
while the solid lines are either $t-$ or $b$ quarks.	}
\label{fg:gghh}
\end{center}
\end{figure}

\begin{figure}[tb]
%\vskip -1in.
\begin{center}
%\hskip -.5in
\includegraphics[scale=0.6]{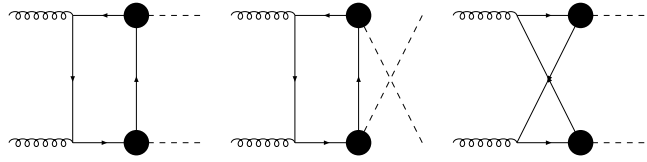}
%\vskip -7. in
\caption{Box contributions to $gg\rightarrow hh$.  The dark circles represent the non-Standard Model contributions, 
while the solid lines are either $t-$ or $b$ quarks.	The crossed diagrams from the initial state are not shown.}
\label{fg:gghhbox}
\end{center}
\end{figure}

Double Higgs production can be analyzed in a similar fashion.  The diagrams shown in Fig.~\ref{fg:gghh}
contribute only to the spin-0 projection, while the box diagrams shown in Fig~\ref{fg:gghhbox}  have both spin-0 and spin-2 
components.
The amplitude for $g^{A,\mu}(p_1)g^{B,\nu}(p_2)\rightarrow h(p_3)h(p_4)$ is 
\begin{equation}
	A^{\mu\nu}_{AB}={\alpha_s\over 3 \pi v^2}\delta_{AB}
	\sum_i
	\biggl[P_1^{\mu\nu}(p_1,p_2)F_0^i(s,t,u,M_j)+
	P_2^{\mu\nu}(p_1,p_2,p_3)F_2^i(s,t,u,M_j)\biggr]\; ,
\label{eq:amp}
\end{equation}
where the sum is over the diagrams, $M_j$ denotes all relevant quark masses, 
$P_1$ and $P_2$ are the orthogonal  projectors onto the spin-$0$ and spin-$2$ states respectively,
\begin{eqnarray}
	P_1^{\mu\nu}(p_1,p_2)&=&
	p_1\cdot p_2 g^{\mu\nu}-p_1^\nu p_2^\mu \; , 
	\nonumber \\
	P_2^{\mu\nu}(p_1,p_2,p_3)&=&
	p_1\cdot p_2 g^{\mu\nu}+{1\over  p_T^2 } \left(
		M_h^2 p_1^\nu p_2^\mu 
		- 2 p_1.p_3 \, p_2^\mu p_3^\nu - 2 p_2.p_3 \, p_1^\nu p_3^\mu + s \, p_3^\mu p_3^\nu
%	+(t-M_H^2) q^\mu p^{\prime \nu} +(u-M_H^2) p^\nu p^{\prime \mu} + s p^{\prime \mu} p^{\prime \nu}
\right)\, ,
\end{eqnarray}
$s,t$, and $u$ are the partonic Mandelstam variables,
\begin{equation}
	s = (p_1+p_2)^2 \; , \quad
	t = (p_1-p_3)^2 \; , \quad
	u = (p_2-p_3)^2 \; ,
\end{equation}
and
$p_T$ is the transverse momentum of the Higgs particle,
\beq 
	p_T^2={ut-M_h^4\over s} \, .
\eeq
The individual  contributions from the diagrams of Figs. \ref{fg:gghh}
and \ref{fg:gghhbox}  to ${\cal O }({1\over M^2_X})$ are:
\begin{eqnarray}
F_0^{(a)}&=& {9 M_h^2\over 4(s-M_h^2)}
\biggl[ (1+\delta_t) F_{1/2}\biggl({4M_{T_1}^2\over s}\biggr)+
(1+\delta_b) F_{1/2}\biggl({4M_{B_1}^2\over s}\biggr)\biggr]\nonumber \\
F_0^{(b)}&=&{9 \over 4 }
\delta_t 
F_{1/2}\biggl({4M_{T_1}^2\over s}\biggr)+
{9 \over 4 }
\delta_b
F_{1/2}\biggl({4M_{B_1}^2\over s}\biggr)
\nonumber \\
F_0^{(c)}&=& -c_{gg}\nonumber \\
F_0^{(d)}&=& -c_g{3M_h^2\over s-M_h^2}\nonumber \\
F_0^{(box)}&=&(1+2\delta_t)F_0^{(box, SM)}(s,t,u,M_{T_1}) +(1+2\delta_b)F_0^{(box,SM)}(s,t,u,M_{B_1})
\label{eftlim}
\end{eqnarray}
where $F_0^{box,SM}(s,t,u,M_{T_1})\rightarrow 1$ for $M_{T_1}\rightarrow \infty$ and $F_0^{(box, SM)}(s,t,u,M_j)$ contains
the $6$ box diagrams with a fermion of mass $M_j$ in the loop.  Analytic results can be found in Refs.
\cite{Plehn:1996wb,Glover:1987nx}\footnote{Our normalization is ${3\over 4}$ times that of 
Ref. \cite{Plehn:1996wb} for the boxes.}.
In the effective theory, the spin-0 contribution is,
\begin{eqnarray}
F_0&=& F_0^{(a)}+ F_0^{(b)}+
 F_0^{(c)}+
 F_0^{(b)}+F_0^{(box)}\nonumber \\&\rightarrow &
 \biggl[ 1-\delta_t -c_{gg}\biggr]-{3M_h^2\over s-M_h^2}\biggl[
 1+\delta_t+c_g\biggr]\label{eftgghh}
 \end{eqnarray}
 where the $2^{nd}$ line is found in the limit $M_{T_1}^2\gg s$ and neglects the $b$ contribution.  
 Taking $c_{gg}=-c_g$,
 \begin{equation}
 F_0\rightarrow \biggl[1+\delta_t+c_g\biggr]F_0^{SM}(M_{T_1}\rightarrow\infty)-2(c_g+\delta_t)\, .
 \end{equation}
The $c_g$ contribution is  in agreement with the LET result of Eq. \ref{rhlet}, while the $\delta_t$
contribution is no longer proportion to the Standard Model result.

 \begin{figure}[tb]
\subfigure[]{
      \includegraphics[width=0.6\textwidth,angle=0,clip]{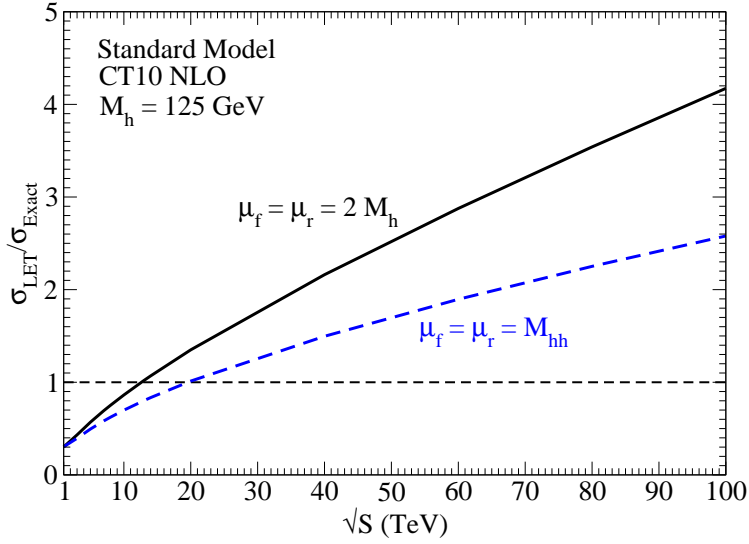}
}
\caption{ Standard Model rate for $pp\rightarrow hh$ from gluon fusion using the LET of Eq. \ref{rhlet}
normalized to the exact cross section.  This plot uses CT10NLO PDFs.}
\label{sm_fig}
\end{figure}

The LET prediction for the  total cross section for double Higgs production in the Standard Model
normalized to the exact result is given in Fig. 
\ref{sm_fig} as a function of center-of-mass energy.   At $\sqrt{S}=13$~TeV, the LET is a reasonable
approximation to the total rate, while at higher energies the deviation from the exact result becomes large. We show this for two choices of factorization and renormalization scales, $\mu_f=\mu_r=2 M_h$ (solid) and $\mu_f=\mu_r=M_{hh}$ (dashed).  The size of the deviation between the LET and exact calculation is very sensitive to the scale choices.

\begin{figure}[tb]
      \includegraphics[width=0.6\textwidth,angle=0,clip]{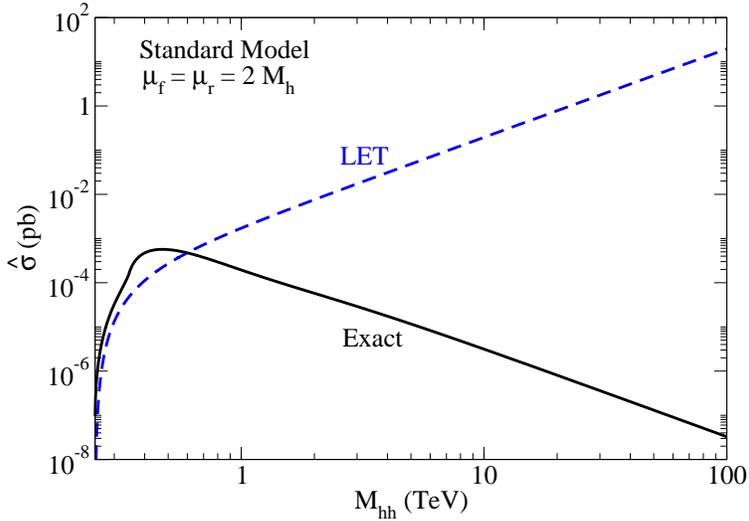}
\caption{ Standard model  partonic cross section for $gg\rightarrow hh$.  }
\label{sm_parton}
\end{figure}

The divergence of the LET from the exact result can be understood by examining the partonic cross section for $gg\rightarrow hh$ shown in Fig.
\ref{sm_parton}.  For partonic sub-energies above around $1$~TeV, the LET and the exact results 
increasingly differ.  The LET contains terms $\sim {M_{hh}^2\over M_{T_1}^2}$, which are not present in 
the exact result.  

\begin{figure}[t]
      \includegraphics[width=0.6\textwidth,angle=0,clip]{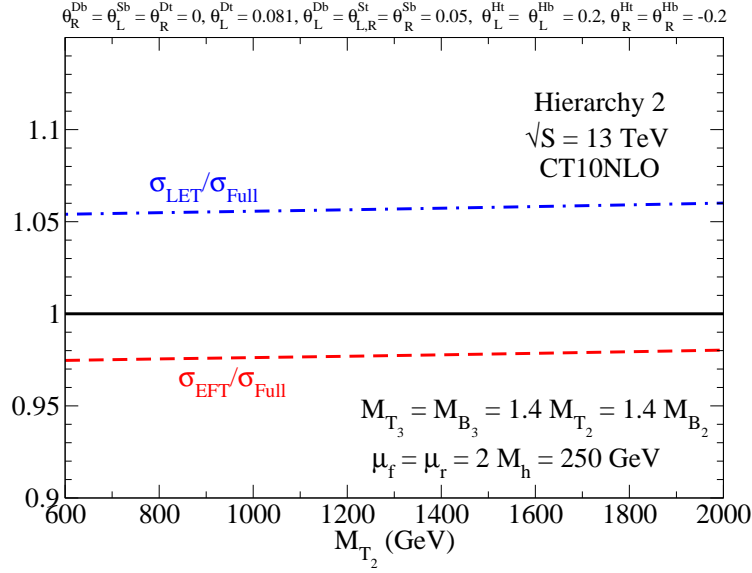}
\caption{ Total cross section for $pp\rightarrow hh$ for a choice of small angles using the
hierarchy  of Section~\ref{hier2sec}.   The EFT and LET results are normalized to the exact one-loop calculation.
}
\label{hier2_fig}
\end{figure}

\begin{figure}[t]
      \includegraphics[width=0.6\textwidth,angle=0,clip]{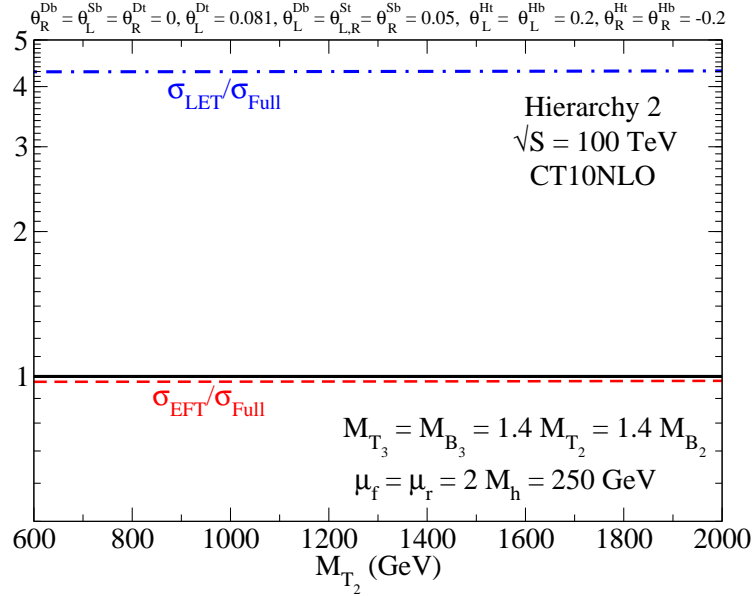}
\caption{ Same as Fig. \ref{hier2_fig}, except $\sqrt{S}=100$~TeV.
}
\label{hier2_figb}
\end{figure}

The first hierarchy of small angles of Section \ref{smallmat} reduces to the Standard Model, so we do not
expect to gain insight from examining this limit.  The second hierarchy, (Section \ref{hier2sec}), however,
is more interesting.  In Figs. \ref{hier2_fig} and \ref{hier2_figb} we show the total
cross sections for $gg\rightarrow hh$ at $\sqrt{S}=13$~TeV and $100$~TeV as
a function of the lightest top partner mass, $M_{T_2}$, for a specific choice of small angles using
the parameterization of Eq. \ref{hier2mat}.
The LET significantly overestimates the rate at $\sqrt{S}=100$~TeV, but is a reasonable
approximation at $\sqrt{S}=13$~TeV.  The EFT, which contains the top and bottom quark contributions exactly,  agrees within
a few percent with the exact calculation.  From Eqs.~\ref{hier2EFT} and \ref{cgHier2}, we see that the EFT and LET depend on differences between the heavy vector-like quark masses and not the overall mass scale.  This result is confirmed in Figs. \ref{hier2_fig} and \ref{hier2_figb}, which show all the results are insensitive to the heavy quark mass scale. 

\begin{figure}[t]
\includegraphics[width=0.66\textwidth,angle=0,clip]{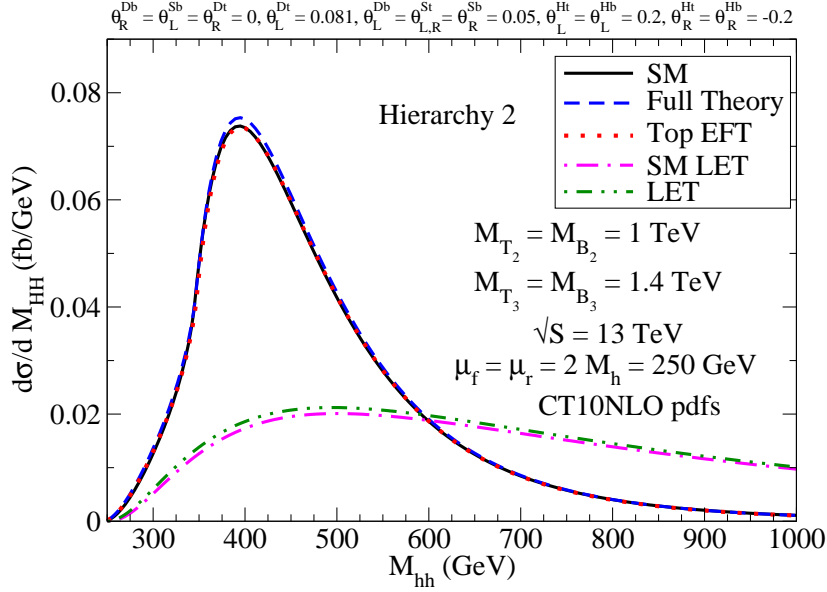}
\caption{ Invariant mass distributions for $pp\rightarrow hh$ at the LHC.  The SM
and SM LET curves represent the exact Standard Model calculation, along with the
LET limit.  The curves labelled Full Theory, Top EFT, and LET are the top partner 
model in the small angle hierarchy of Section~\ref{hier2sec}, using the exact one-loop
calculation, the EFT of Eq. \ref{eftlim}, and the LET of Eq. \ref{rhlet}.
}
\label{hier2dif_fig}
\end{figure}

\begin{figure}[t]
\includegraphics[width=0.66\textwidth,angle=0,clip]{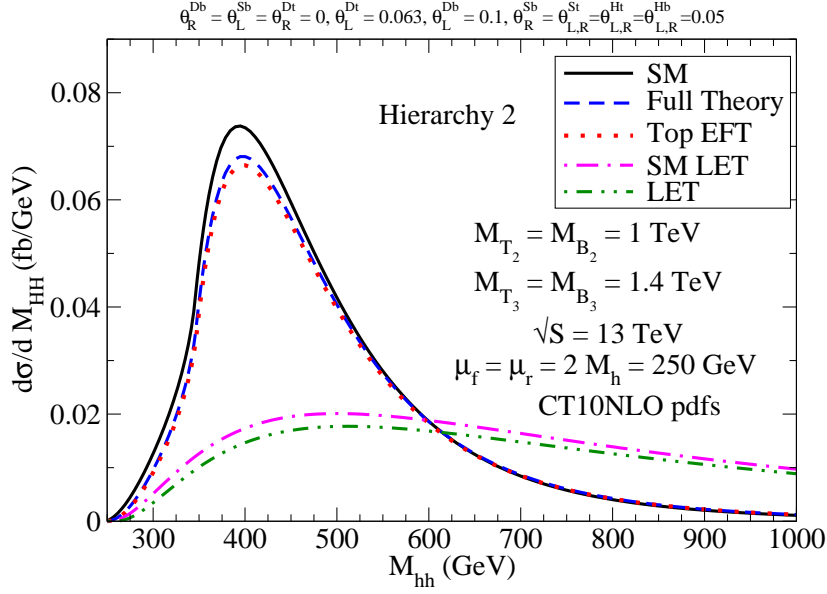}
\caption{ Same as Fig.~\ref{hier2dif_fig} with a different parameter point.
}
\label{hier2dif_fig1}
\end{figure}

\begin{figure}[t]
      \includegraphics[width=0.66\textwidth,angle=0,clip]{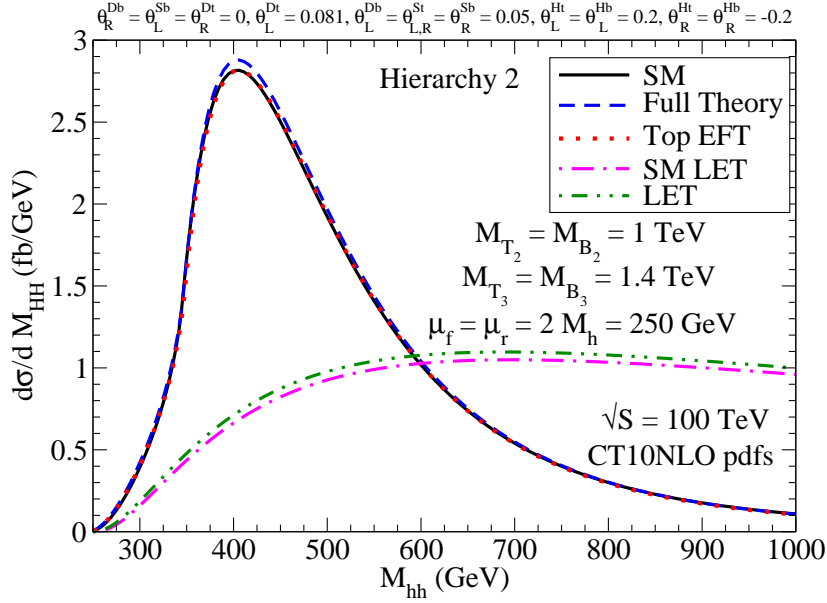}
\caption{ Same as Fig.~\ref{hier2dif_fig} with $\sqrt{S}=100$~TeV.
}
\label{hier2dif_fig100}
\end{figure}

\begin{figure}[t]
      \includegraphics[width=0.66\textwidth,angle=0,clip]{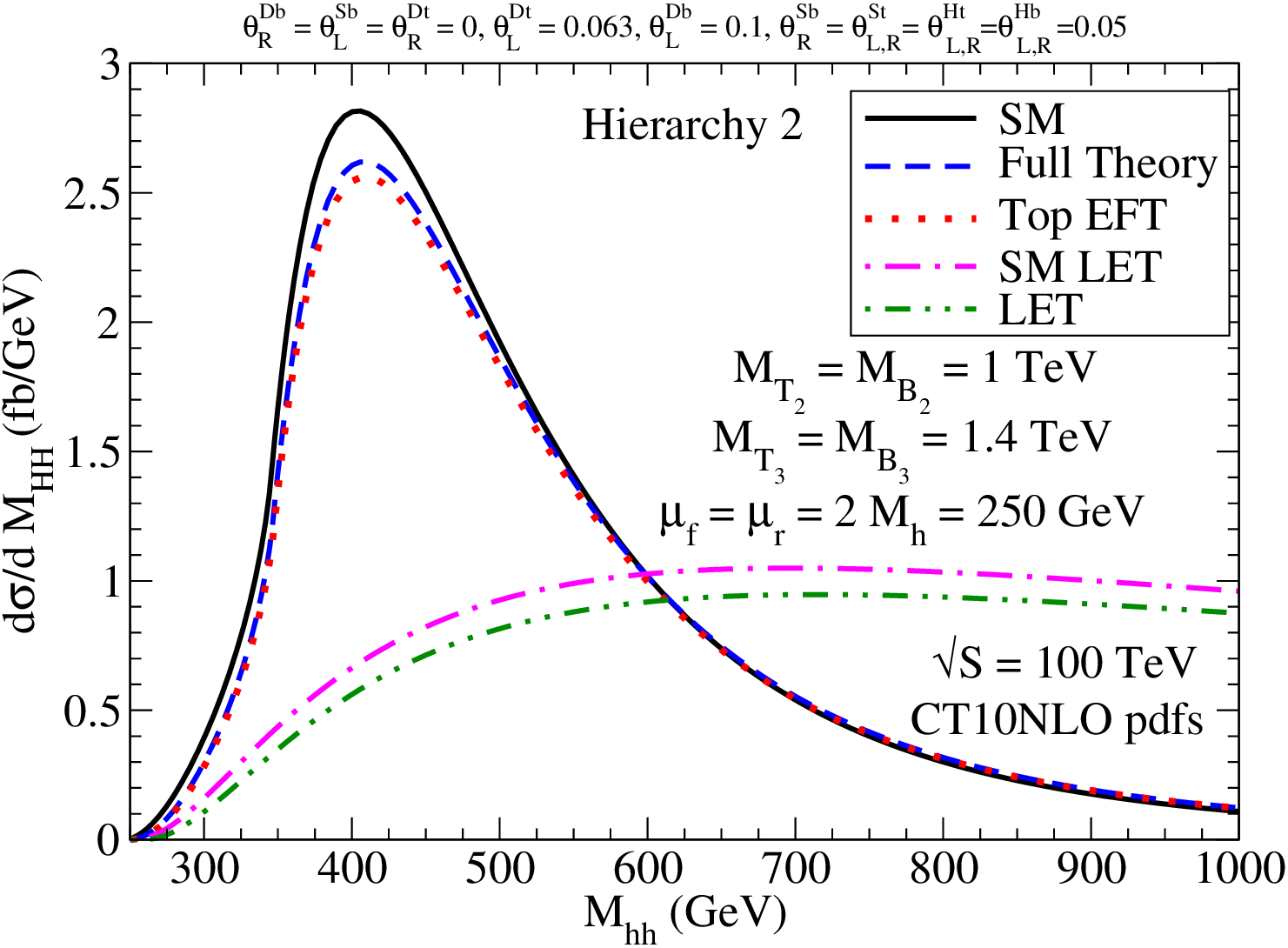}
\caption{ Same as Fig.~\ref{hier2dif_fig1} with $\sqrt{S}=100$~TeV.
}
\label{hier2dif_fig1001}
\end{figure}

It is well known that the LET does not accurately reproduce distributions for
double Higgs production\cite{Dawson:2012mk,Gillioz:2012se,Baur:2002rb}. For a choice of small angles and heavy quark masses, we
show the invariant mass distribution of the Higgs bosons, $d\sigma\over dM_{hh}$, in Figs.~\ref{hier2dif_fig}, \ref{hier2dif_fig1},  \ref{hier2dif_fig100}, and \ref{hier2dif_fig1001} at the LHC with $\sqrt{S}=13$ and $100$~TeV.  We include the Standard Model distributions for comparison.  The LET
does a poor job of reproducing the exact distributions, both in the Standard
Model and in the top partner model.  The curves labelled ``SM'' and ``Full Theory'' contain the
exact one-loop calculations for the Standard Model and top partner model respectively, while the curve labelled ``Top EFT'' is the top partner model calculation using
the results of Eq. \ref{effective}.  The EFT reproduces the exact calculation quite accurately.  We show this for two parameter points to illustrate the robustness of this conclusion.  Both points reproduce the Standard Model single Higgs production rate to within $\sim10\%$.  In a given model,
therefore, the EFT can be used not only for the total rate, but also for distributions.   The distributions in the top
partner model are quite similar to the Standard Model.  Scanning over small angles, we were not able to
find an example with a large deviation from the Standard Model. 

\section{Conclusions}

We considered a scenario with both $SU(2)_L$ singlet and doublet vector-like fermions.  Such a 
scenario could in principle have large deviations from the Standard Model predictions for single
 and double
Higgs production.  However, we were unable to find parameters consistent with electroweak
precision measurements and the single Higgs production rate  which gave a significant deviation
from the Standard Model prediction for double Higgs production.

We constructed two versions of an effective theory.  The well known low energy theorem
(LET) treats all fermions
as infinitely massive.  The total cross section for Higgs pair production
is well approximated by the LET at $\sqrt{S}=13$~TeV, but increasingly differs at higher energies.  The LET cannot reproduce the invariant
mass distribution of the $hh$ pairs.  In order to include top quark mass effects, we derived an effective
Lagrangian (EFT) containing only light fermions, but with non-Standard Model coefficients, which we computed to
${\cal O}({1\over M^2_X})$.  The EFT obtains accurate results for both total and differential double Higgs
rates.  Our results can be used to reliably compute the leading effects of models with heavy vector-like fermions. 

An important result is
the observation that the coefficients of the effective Lagrangian of Eq. 
\ref{effective} are not free parameters, but are related to each other in any consistent model.  Despite
the proliferation of Yukawa couplings in Eq. \ref{massdef}, a consistent treatment
yields an effective Lagrangian which depends on only $3$ parameters, $\delta_b$, $\delta_t$, and $c_g$.  This is similar to the case in composite Higgs models where deviations in Yukawa couplings and new effective operators relevant for double Higgs production are tightly correlated~\cite{Grober:2010yv}.  Hence, we expect the EFT used to study Higgs production in composite Higgs models to be a very good approximation to a complete calculation.
\label{concsec}

\section*{Acknowledgements}
This work is supported by the U.S. Department of Energy under grant
No.~DE-AC02-98CH10886. 
\newpage

\appendix
\section*{Appendix A:  Oblique Parameters}
The limits on the parameters of
the fermion sector arising from contributions to  gauge
boson 2-point functions  can
be studied using the $S$,$T$ and $U$ functions
following the notation of Peskin and
Takeuchi\cite{Peskin:1991sw},
\begin{eqnarray}
\alpha S&=&
\biggl({4 s_W^2 c_W^2\over M_Z^2}\biggr)
\biggl\{ \Pi_{ZZ}(M_Z^2)- \Pi_{ZZ}(0)-
\Pi_{\gamma\gamma}(M_Z^2)
-{c_W^2-s_W^2\over c_W s_W}
\Pi_{\gamma Z}(M_Z^2)
\biggr\}\nonumber \\
\alpha  T &=&
 \biggl({ \Pi_{WW}(0)\over M_W^2}-{
\Pi_{ZZ}(0)\over M_Z^2}
\biggr)\, .
\label{sdef}
\end{eqnarray}
In terms of the mixing angles and the mass eigenstates of the full theory, the contributions from
heavy quarks, including the Standard Model top and bottom quarks, 
to $\Delta T$ and $\Delta S$ 
are\cite{Lavoura:1992np,He:2001tp},\footnote{ We assume all entries in the mixing matrices are real.}
\begin{eqnarray}
\Delta T&=& {N_c\over 16\pi s_W^2 M_W^2}
\biggl\{ \Sigma_{i,j=1,2,3} 
\biggl[
\biggl( \mid U_{L,ij}\mid^2+\mid U_{R,ij}\mid^2\biggr)
\theta_+(M_{T_i},M_{B_j})
+2U_{L,ij}U_{R,ij}^\dagger\theta_-(M_{T_i},M_{B_j})\biggr]\nonumber \\
&&
-\Sigma_{i<j=1,2,3} 
\biggl[
\biggl( 
\mid X_{L,ij}^t\mid^2
+\mid X_{R,ij}^t \mid^2\biggr)
\theta_+(M_{T_i},M_{T_j})
+2X_{L,ij}^t X_{R,ij}^{t\dagger}\theta_-(M_{T_i},M_{T_j})\biggr]\nonumber \\
&&
-\Sigma_{i<j=1,2,3} 
\biggl[
\biggl( 
\mid X_{L,ij}^b\mid^2
+\mid X_{R,ij}^b\mid^2\biggr)
\theta_+(M_{B_i},M_{B_j})
+2 X_{L,ij} ^bX_{R,ij}^{b\dagger}\theta_-(M_{B_i},M_{B_j})\biggr]
\biggr\}\nonumber \\
\Delta S&=& {N_c\over 2\pi M_Z^2 }
\biggl\{ \Sigma_{i,j=1,2,3} 
\biggl[
\biggl( \mid U_{L,ij}\mid^2+\mid U_{R,ij}\mid^2\biggr)
\psi_+(M_{T_i},M_{B_j})
+2U_{L,ij}U_{R,ij}^\dagger\psi_-(M_{T_i},M_{B_j})\biggr]\nonumber \\
&&
-\Sigma_{i<j=1,2,3} 
\biggl[
\biggl( 
\mid X_{L,ij}^t\mid^2
+\mid X_{R,ij}^t \mid^2\biggr)
\chi_+(M_{T_i},M_{T_j})
+2X_{L,ij}^t X_{R,ij}^{t\dagger}\chi_-(M_{T_i},M_{T_j})\biggr]\nonumber \\
&&
-\Sigma_{i<j=1,2,3} 
\biggl[
\biggl( 
\mid X_{L,ij}^b\mid^2
+\mid X_{R,ij}^b\mid^2\biggr)
\chi_+(M_{B_i},M_{B_j})
+2 X_{L,ij} ^bX_{R,ij}^{b\dagger}\chi_-(M_{B_i},M_{B_j})\biggr]
\biggr\}\, ,
\label{genstu}
\end{eqnarray}
where the functions $\theta_\pm, \chi_\pm$ are defined below and $N_c=3$.

\begin{eqnarray}
\theta_+(m_1,m_2)&=& 
m_1^2+m_2^2-{2m_1^2m_2^2\over m_1^2-m_2^2}\ln\biggl({m_1^2\over m_2^2}\biggr)\nonumber \\
\theta_-(m_1,m_2)&=&
 2m_1 m_2 \biggl[{m_1^2+m_2^2\over m_1^2-m_2^2}\ln\biggl({m_1^2\over m_2^2}\biggr)-2\biggr]
\nonumber \\ \theta_+(m,m)&=& 0\nonumber \\
\theta_-(m,m)&=& 0\nonumber \\
\end{eqnarray}
and
\begin{eqnarray}
\psi_+(m_1,m_2)&=& 
{22m_1^2+14m_2^2\over 9}
-{M_Z^2\over 9}\log\biggl({m_1^2\over m_2^2}\biggr)
+{11m_1^2+M_Z^2\over 18} f(m_1,m_1)
\nonumber \\ &&
+{7m_1^2-M_Z^2\over 18}f(m_2,m_2)
\nonumber \\
\psi_-(m_1,m_2)&=& -\mid
(m_1 m_2 \mid\biggl[4+{1\over 2}\biggl(f(m_1,m_1)+f(m_2,m_2)\biggr)\biggr]\nonumber \\
\chi_+(m_1,m_2)&=&
 {m_1^2+m_2^2\over 2}-{(m_1^2-m_2^2)^2\over 3 M_Z^2}
+\biggl[
{(m_1^2-m_2^2)^3\over 6M_Z^4}
-\biggl({M_Z^2\over 2}\biggr){m_1^2+m_2^2\over m_1^2-m_2^2}\biggr]
\ln\biggl(
{m_1^2\over m_2^2}\biggr)
\nonumber \\
&&+{m_1^2-M_Z^2\over 6}f(m_1,m_2)+{m_2^2-M_Z^2\over 6}f(m_2,m_2)
\nonumber \\ &&
+\biggl[{M_Z^2\over 3} -{m_1^2+m_2^2\over 6}-{(m_1^2-m_2^2)^2\over 6 M_Z^2}\biggr]f(m_1,m_2)
\nonumber \\
\chi_-(m_1,m_2)&=& -\mid m_1 m_2\mid \biggl[2 +\biggl(
{m_1^2-m_2^2\over M_Z^2}
-{m_1^2+m_2^2\over m_1^2-m_2^2}\biggr)\ln\biggl({m_1^2\over m_2^2}
\biggr)\nonumber \\
&&+{1\over 2}\biggl(f(m_1,m_1)+f(m_2,m_2)\biggr)-f(m_1,m_2)\biggr]
\nonumber \\
\chi_+(m,m)&=&0\nonumber \\
\chi_-(m,m)&=&0
\end{eqnarray}
and
\begin{eqnarray}
f(m_1,m_2)&= &-\biggl({2{\sqrt{\Delta}\over M_Z}\biggl)
\bigg[
\arctan\biggl({m_1^2-m_2^2+M_Z^2\over M_Z \sqrt{\Delta}}\biggr)
-\arctan\biggl({m_1^2-m_2^2-M_Z^2\over M_Z\sqrt{\Delta}}\biggr)\biggr]
\quad {\hbox{if}}}~\, \Delta > 0
\nonumber \\
&= &0
 \qquad {\hbox{if}}~\,, \Delta=0\nonumber \\
&=&{1\over M_Z}\sqrt{-\Delta}\ln\biggl({m_1^2+m_2^2-M_Z^2+M_Z\sqrt{-\Delta}
\over m_1^2+m_2^2-M_Z^2-M_Z\sqrt{-\Delta} }\biggr) \qquad \hbox{if}~\, \Delta < 0
\nonumber \\
\Delta&=&-M_Z^2-{m_1^4+m_2^4\over M_Z^2}
+2m_1^2+2m_2^2+{2m_1^2 m_2^2\over M_Z^2}\nonumber \\
&=&-M_Z^2\biggl(1-{m_1^2+m_2^2\over M_Z^2}\biggr)^2+{4m_1^2m_2^2\over M_Z^2}
\end{eqnarray}
\newpage
\bibliographystyle{hunsrt}
\bibliography{fermions}

\end{document}